\documentclass{article}
\usepackage{emulateapj,pstricks,apjfonts}

\def\asca       {{\em ASCA}\/}
\def\rosat      {{\em ROSAT}\/}
\def\chandra    {{\em Chandra}\/}
\def\xmm        {{\em XMM}\/}
\def\sax        {{\em SAX}\/}

\def\cmsq       {cm$^{-2}$}

\def\ergscmam   {erg$\;$s$^{-1}\,$cm$^{-2}\,$arcmin$^{-2}$}

\def\ergscm     {erg$\;$s$^{-1}\,$cm$^{-2}$}

\def\ctssam     {cts$\;$s$^{-1}\,$arcmin$^{-2}$}

\def\phscmam    {phot$\;$s$^{-1}\,$cm$^{-2}\,$arcmin$^{-2}$}

\def\phcmskevsr {phot$\;$s$^{-1}\,$cm$^{-2}\,$keV$^{-1}\,$sr$^{-1}$}

\def\gax        {\gtrsim}
\def\lax        {\lesssim}

\def\bi{\bfseries\itshape}

\begin{document}

\submitted{ApJ in press; astro-ph/0209441}

\lefthead{DIFFUSE X-RAY BACKGROUND}
\righthead{MARKEVITCH ET AL.}

\title{{\em CHANDRA}\/ SPECTRA OF THE SOFT X-RAY DIFFUSE BACKGROUND}

\author{M.~Markevitch, M.~W.~Bautz$^1$, B.~Biller, Y.~Butt, R.~Edgar,
T.~Gaetz, G.~Garmire$^2$, C.~E.~Grant$^1$, P.~Green, M.~Juda,
P.~P.~Plucinsky, D.~Schwartz, R.~Smith, A.~Vikhlinin, S.~Virani,
B.~J.~Wargelin, S.~Wolk}

\affil{Harvard-Smithsonian Center for Astrophysics, 60 Garden St.,
Cambridge, MA 02138; maxim@head-cfa.harvard.edu}

\footnotetext[1]{MIT ~~$^2$ PSU}
\setcounter{footnote}{2}

\begin{abstract}

We present an exploratory \chandra\ ACIS-S3 study of the diffuse component
of the Cosmic X-ray Background in the 0.3--7 keV band for four directions at
high Galactic latitudes, with emphasis on details of the ACIS instrumental
background modeling.  Observations of the dark Moon are used to model the
detector background. A comparison of the Moon data and the data obtained
with ACIS stowed outside the focal area showed that the dark Moon does not
emit significantly in our band. Point sources down to $3\times 10^{-16}$
\ergscm\ in the 0.5--2 keV band are excluded in our two deepest
observations. We estimate the contribution of fainter, undetected sources to
be less than 20\% of the remaining CXB flux in this band in all four
pointings. In the 0.3--1 keV band, the diffuse signal varies strongly from
field to field and contributes between 55\% and 90\% of the total CXB
signal. It is dominated by emission lines that can be modeled by a
$kT=0.1-0.4$ keV plasma. In particular, the two fields located away from
bright Galactic features show a prominent line blend at $E\approx 580$ eV
(O{\small VII} + O{\small VIII}) and a possible line feature at $E\sim 300$
eV. The two pointings toward the North Polar Spur exhibit a brighter O blend
and additional bright lines at 730--830 eV (Fe{\small XVII}). We measure the
total 1--2 keV flux of $(1.0-1.2\,\pm0.2)\times 10^{-15}$ \ergscmam\ (mostly
resolved), and the 2--7 keV flux of $(4.0-4.5\,\pm1.5)\times 10^{-15}$
\ergscmam.  At $E>2$ keV, the diffuse emission is consistent with zero, to
an accuracy limited by the short Moon exposure and systematic uncertainties
of the S3 background.  Assuming Galactic or local origin of the line
emission, we put an upper limit of $\sim 3 \times 10^{-15}$ \ergscmam\ on
the 0.3--1 keV extragalactic diffuse flux.

\end{abstract}


\section{INTRODUCTION}

The existence of a cosmic X-ray background (CXB) was one of the first
discoveries of extra-solar X-ray astronomy (Giacconi et al.\ 1962).  In the
intervening four decades, observations with improving angular and spectral
resolution have enhanced our understanding of the components that make up
this background.  Several broad-band, all-sky surveys have been performed
using proportional-counter detectors (Marshall et al.\ 1980; McCammon et
al.\ 1983; Marshall \& Clark 1984; Garmire et al.\ 1992; Snowden et al.\
1995, 1997; for a review of pre-\rosat\ results, see McCammon \& Sanders
1990). These surveys form a consistent picture of the angular distribution
of X-ray emission in the various bands.  Above 2 keV, the emission is highly
isotropic on large angular scales and has an extragalactic origin.  Below 2
keV, the X-ray background is a mixture of Galactic diffuse emission (e.g.,
Kuntz \& Snowden 2000 and references therein), heliospheric and geocoronal
diffuse component (e.g., Cravens 2000), and extragalactic flux from point
sources and, possibly, from intergalactic warm gas that may contain the bulk
of the present-day baryons (e.g., Cen \& Ostriker 1999).

The earliest observations could provide only limited spectral information on
the background.  Marshall et al.\ (1980) found that the spectrum in the
3--50 keV range was well fit by a thermal bremsstrahlung model with $kT \sim
40$ keV.  In the 3--10 keV band this can be approximated by a power law with
a photon index of $-1.4$. At energies below 1 kev the background surface
brightness exceeds the extrapolation of this power law (Bunner et al.\
1969).  Later observations with gas-scintillation proportional counters and
solid-state detectors (Inoue et al.\ 1979; Schnopper et al.\ 1982; Rocchia
et al.\ 1984) suggested emission lines in the 0.5--1.0 keV band, most likely
from oxygen.  The evidence for emission lines in this band has become more
convincing in recent observations using CCDs (Gendreau et al.\ 1995;
Mendenhall \& Burrows 2001) and in a high-resolution spectrum obtained by
McCammon et al.\ (2002) in a microcalorimetric experiment.  A definitive
demonstration of spectral lines in the 0.15--0.3 keV band (at low Galactic
latitude) was obtained by Sanders et al.\ (2001) using a Bragg-crystal
spectrometer.  Observations that combine high spectral and angular
resolution are essential for disentangling the many CXB soft emission
components.

%%%%%%%%%%%%%%%%%%%%%%%%%%%%%%%%%%%%%%%%%%%%%%%%%%%%%%%%%%%%%%%%%%%%%%%%%%
\begin{table*}[t]
\begin{center}
\begin{minipage}{16.5cm}
\renewcommand{\arraystretch}{1.4}
\renewcommand{\tabcolsep}{2.2mm}
\small
\begin{center}
\caption{Data Summary}
\begin{tabular}{p{5.8cm}ccccc}
\hline 
\hline
Observation Identificator (OBSID)  & 3013 & 3419 & 869    & 930  & Moon \\
\hline
$(l,b)$, deg\dotfill& $(259.6, +56.9)$ & $(187.1, -31.0)$ & $(36.6, +53.0)$ & $(358.7, +64.8)$&...\\
Galactic $N_H$, $10^{20}$\cmsq \dotfill    & 4.1 & 11.4 & 4.3  & 1.8 &... \\
\rosat\ R4--R5 flux, $10^{-6}$\ctssam\dotfill&160 & 90 &200--250$^a$ &400&...\\
Observation date \dotfill & 2001 Dec 13 & 2002 Jan 08 & 2000 Jun 24 &%
                            2000 Apr 19 & 2001 Jul 26\\
Total (uncleaned) exposure, ks      \dotfill & 112 & 98   & 57  & 40 & 16  \\
Exposure for source detection, ks   \dotfill & 101 & 92   & 52  & 28 & ... \\
Exposure for background spectra, ks \dotfill & 69  & 86   & 52  & 20 & 11  \\
Field solid angle, arcmin$^2$       \dotfill & 69  & 69   & 51  & 69 & 70  \\
\hline
\label{table:obslist}
\end{tabular}
\end{center}
\vspace{-6mm}
{\footnotesize
$^a$ Affected by an artifact in the \rosat\ all-sky map
}
\vspace{-2mm}
\end{minipage}
\end{center}
\end{table*}
%%%%%%%%%%%%%%%%%%%%%%%%%%%%%%%%%%%%%%%%%%%%%%%%%%%%%%%%%%%%%%%%%%%%%%%%%%

\chandra\ and \xmm\ should soon provide a wealth of new information on the
CXB. Several works have already taken advantage of the \chandra's arcsecond
resolution to study the point source component of the CXB (e.g., Mushotzky
et al.\ 2000; Brandt et al.\ 2001; Rosati et al.\ 2002). The first \xmm\
results are starting to appear as well (De Luca \& Molendi 2002; Warwick
2002; Lumb et al.\ 2002), utilizing the large effective area of that
observatory.

In this paper, we present a \chandra\ ACIS study of the diffuse CXB at high
Galactic latitudes. The main advantage of \chandra\ over all other
instruments is its ability to resolve point sources down to very low fluxes
and probe the true diffuse background. In addition to that, compared to
\rosat\ PSPC (which had lower detector background), ACIS has energy
resolution sufficient to identify spectral lines. Compared to \xmm\ EPIC,
ACIS appears to be less affected by instrumental background flares (although
during quiescent periods, the ACIS detector background per unit sky signal
is higher). To the extent that results can be compared, we confirm many of
the recent findings made with other instruments.

Technical aspects of our study, especially the ACIS instrumental background
modeling, are quite complex, and we discuss them here in detail. Much of our
analysis procedure may be useful for studies of extended sources such as
clusters of galaxies. Uncertainties are $1\sigma$ unless specified
otherwise.

\section{DATA SET} 

\begin{figure*}[b]
\center
\pspicture(0,18.0)(8.75,24.0)
%\psgrid(0,0)(18.5,24)

\rput[tl]{0}(0.1,23.9){\epsfxsize=8.5cm \epsfclipon
\epsffile{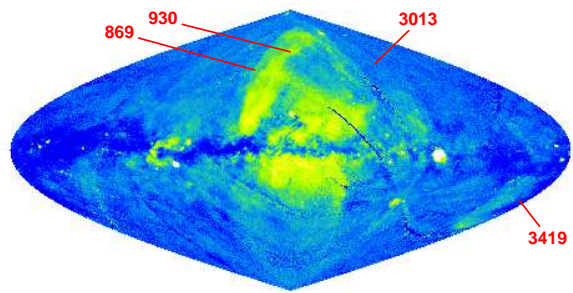}}

\rput[tl]{0}(0.0,19.2){
\begin{minipage}{8.75cm}
\small\parindent=3.5mm

\caption{\rosat\ PSPC all-sky map of CXB in the R4--R5 or 3/4 keV band
  (Snowden et al.\ 1997) in Galactic coordinates. Positions of our
  observations are marked; labels give OBSIDs.}
\label{fig:sxrb}
\par
\end{minipage}
}

\endpspicture
\end{figure*}

For this exploratory study of the diffuse CXB, we selected four \chandra\
ACIS-S observations at high Galactic latitudes, listed in
Table~\ref{table:obslist}.  Our main focus is two relatively deep (90--100
ks) observations 3013 and 3419, obtained at positions away from any bright
Galactic features seen in the \rosat\ All-Sky Survey (RASS) R4-R5 or 3/4 keV
band (Snowden et al.\ 1997).  For comparison, we also analyze two shorter
archival observations towards an edge (OBSID 869) and the middle (OBSID 930)
of the North Polar Spur%
\footnote{The North Polar Spur, part of the Loop I supershell of emission in
the RASS (Egger \& Aschenbach 1995) is thought to be the collision of
explosive remnants with the Local Bubble.}
which exhibits bright emission in the \rosat\ 3/4 keV band. In
Fig.~\ref{fig:sxrb}, positions of the four observations are overlaid on the
RASS 3/4 keV map. None of these observations' original goals were related to
CXB, and there are no nearby cataloged extended sources, except an irregular
galaxy that was a target of OBSID 869, spatially excluded from our analysis.

\section{ACIS INSTRUMENTAL BACKGROUND}
\label{sec:analysis}

A critical part of this study is modeling of the ACIS instrumental
background. This background is caused by cosmic charged particles and
consists of a slowly changing quiescent component, and at least two species
of highly variable background flares whose spectra are very different from
that of the quiescent component. Below we describe in detail how these
components were dealt with from a practical perspective; their exact
physical nature is beyond the scope of this paper.

\subsection{Background flare filtering}
\label{sec:lc}

\begin{figure*}[t]
\pspicture(0,8.6)(18.5,24.0)
%\psgrid(0,0)(18.5,24)

\rput[tl]{0}(0.,24){\epsfxsize=9cm
\epsffile{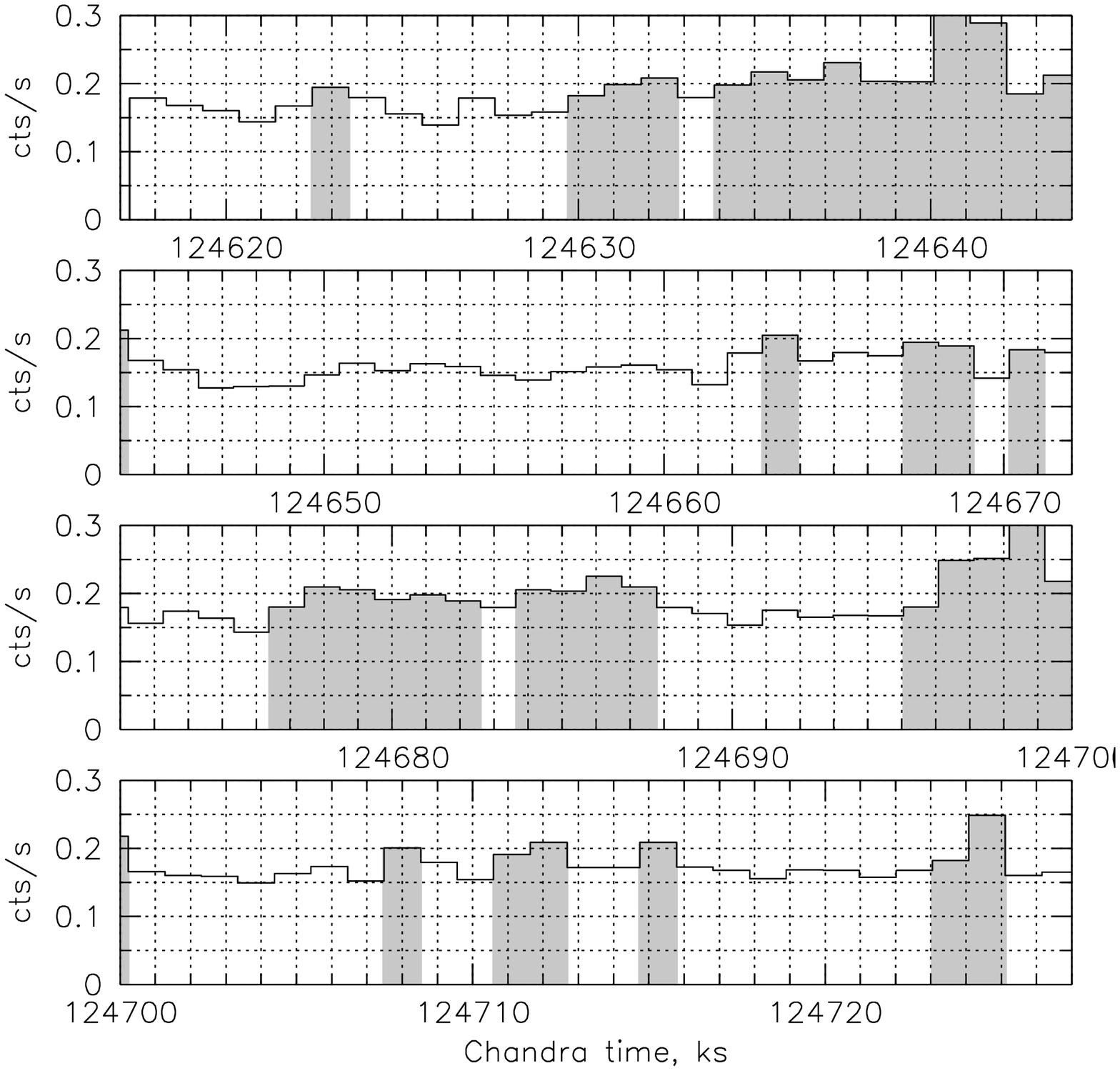}}

\rput[tl]{0}(9.6,24){\epsfxsize=9cm
\epsffile{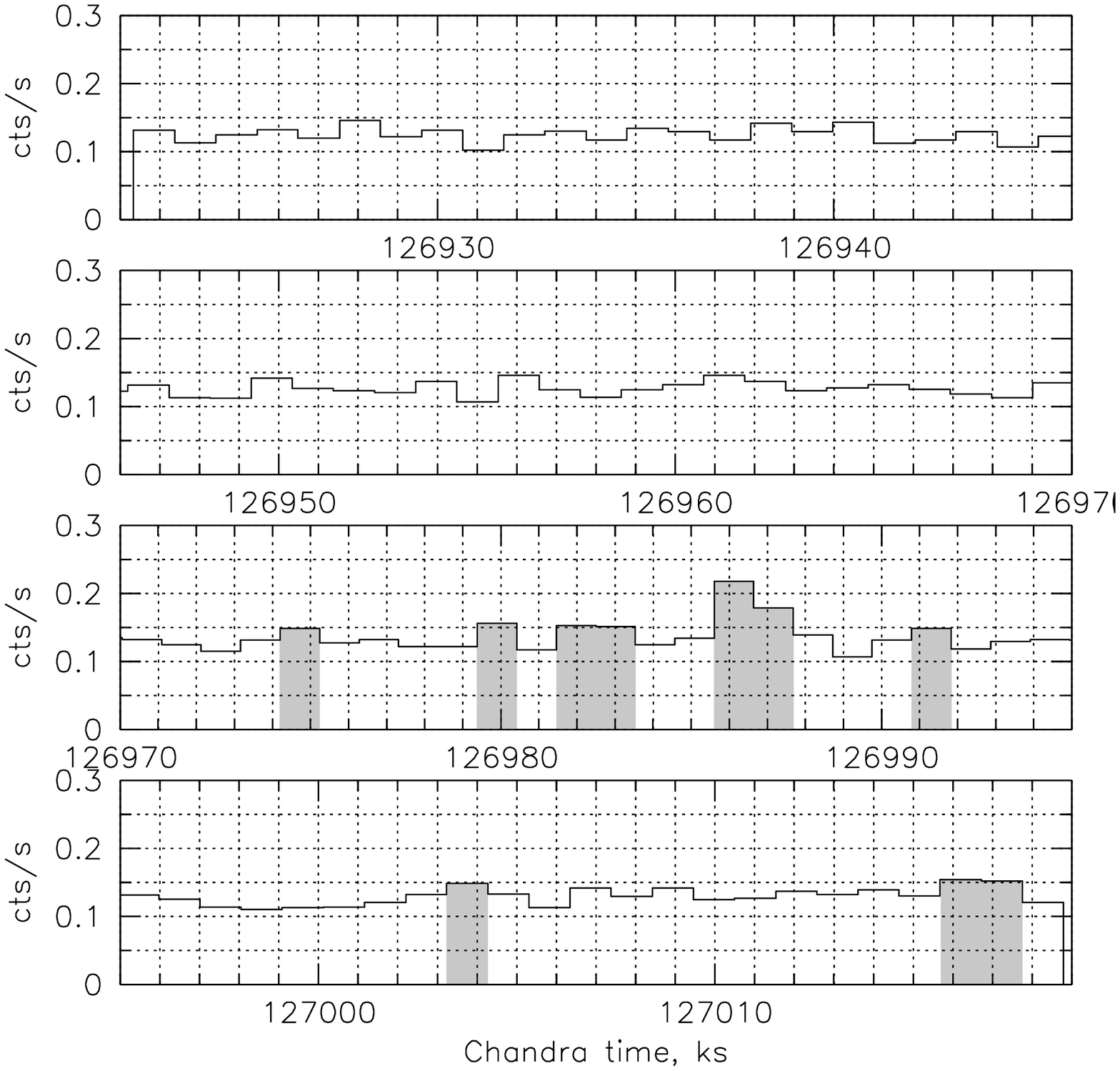}}

\rput[tl]{0}(0.,14.8){\epsfxsize=9cm
\epsffile{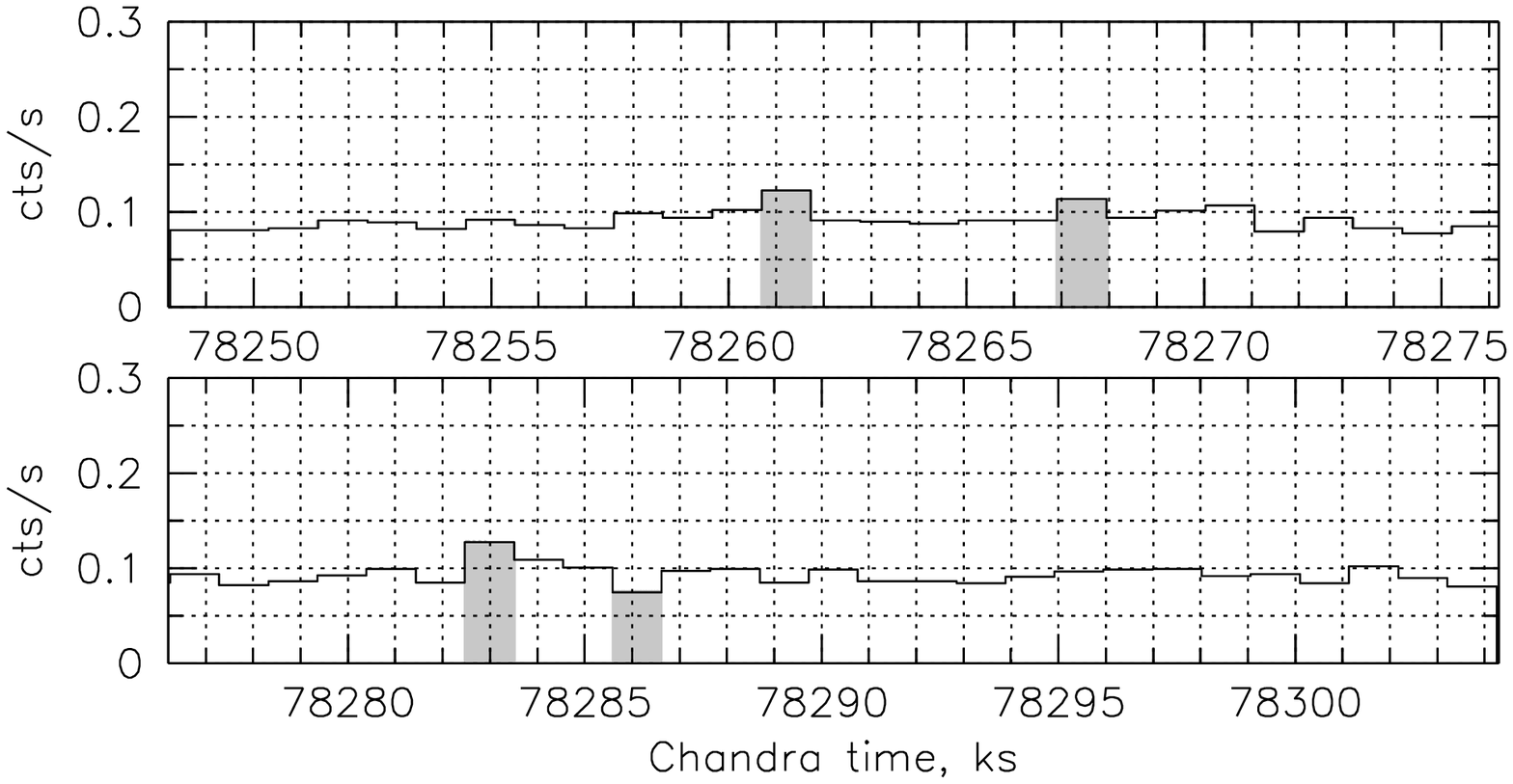}}

\rput[tl]{0}(9.6,14.8){\epsfxsize=9cm
\epsffile{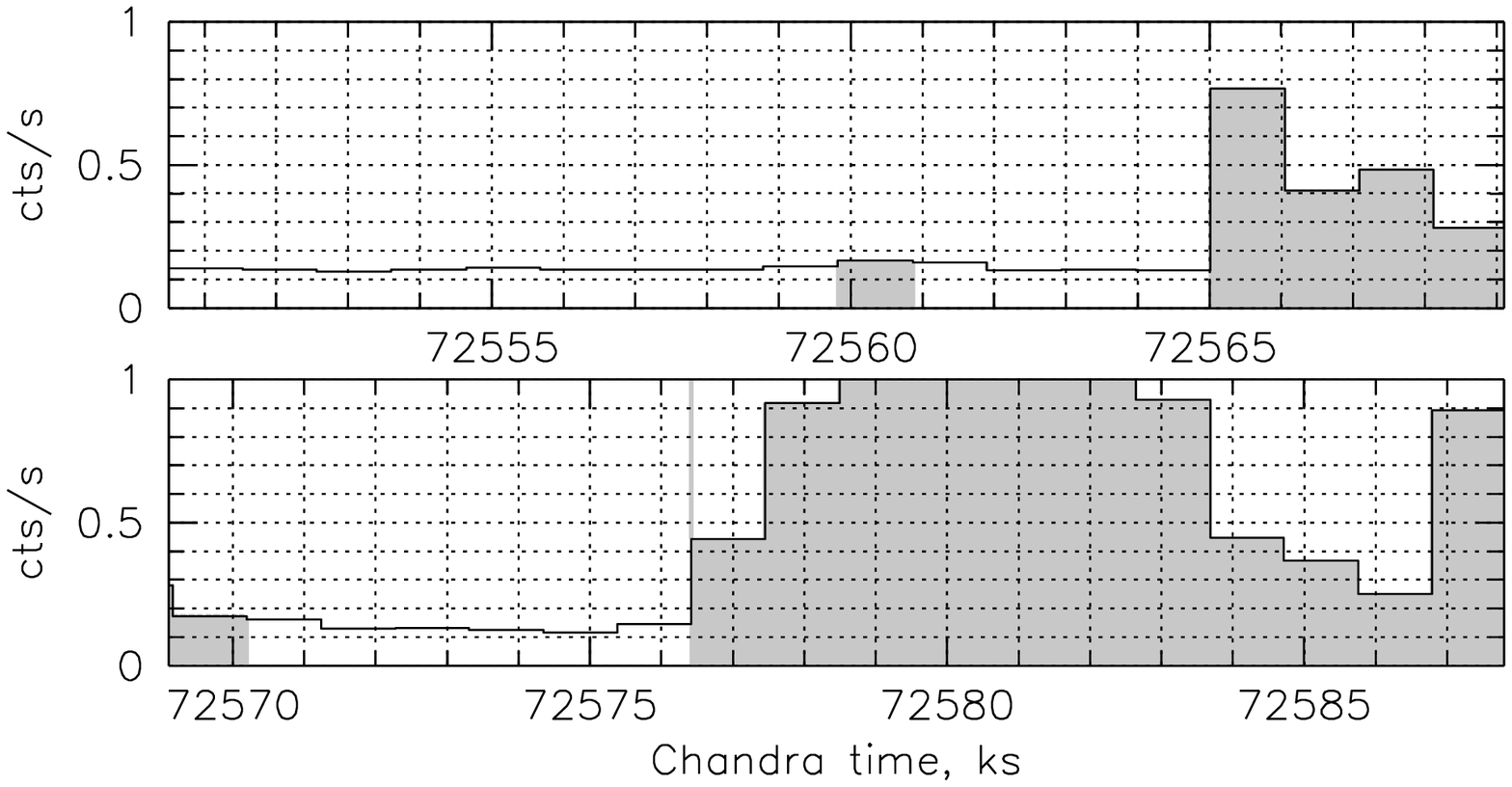}}

\rput[ll]{0}(1.3,23.4){\bf 3013}
\rput[ll]{0}(10.95,23.4){\bf 3419}
\rput[ll]{0}(1.25,14.2){\bf 869}
\rput[ll]{0}(10.9,14.2){\bf 930}

\rput[tl]{0}(0.0,9.6){
\begin{minipage}{18.5cm}
\small\parindent=3.5mm

\caption{Light curves of the four CXB observations in the 2.5--7 keV band,
  where the contribution of flares is most easily detected. OBSIDs are
  marked in each panel. For OBSID 869, only 72\% of the chip area is
  used. VF mode filtering was applied (\S\ref{sec:vf}). Shaded bins are
  above or below a factor of 1.2 of the quiescent level and are excluded.}
\label{fig:lc}
\par
\end{minipage}
}

\endpspicture
\end{figure*}

We use data from the ACIS backside-illuminated (BI) chip S3. Compared to the
ACIS frontside-illuminated (FI) chips, S3 has a higher sensitivity at low
energies. However, its sensitivity to low-energy X-rays also renders it more
sensitive to particle events, which results in more frequent background
flares than in FI chips (Plucinsky \& Virani 2000; Markevitch 2001). The
quiescent background is stable and predictable; therefore, when the accuracy
of the background subtraction is critical, it is best to exclude flare
periods from the analysis. The spectra of the flaring and quiescent
background components (discussed below) are such that the best energy band
in which to look for flares in chip S3 is approximately 2.5--7 keV.
Figure~\ref{fig:lc} shows light curves in this energy band for our four
observations (from the whole S3 chip, excluding celestial sources). The time
bin size ($\sim 1$ ks) is chosen to limit the statistical scatter while
providing a reasonably detailed light curve.

In observations 3419, 869, and 930, the quiescent rate is easily
identifiable and very close to that in most other observations performed
during 2000--2001. To limit the background modeling uncertainty, we exclude
from further analysis all time bins above and below a factor of 1.2 of this
rate (the rate can be lower, for example, because of occasional short
intervals of missing telemetry, bad aspect, etc.) The resulting clean
exposures are given in Table~\ref{table:obslist}.

\subsubsection{Anomalous background in field 3013}
\label{sec:lc3013}

Observation 3013 is unusual in that the apparent ``quiescent'' rate between
the numerous flare intervals (Fig.~\ref{fig:lc}) is about 30\% higher than
in other observations.  It is also more variable than the quiescent rate
usually is.  It appears that, in fact, all of this exposure is affected by a
long flare, which requires special treatment (and unfortunately, will add to
the systematic uncertainty of the results).

\begin{figure*}[t]
\pspicture(0,14.0)(18.5,23.4)
%\psgrid(0,0)(18.5,24)

\rput[tl]{0}(-0.2,24){\epsfxsize=9cm \epsfclipon
\epsffile{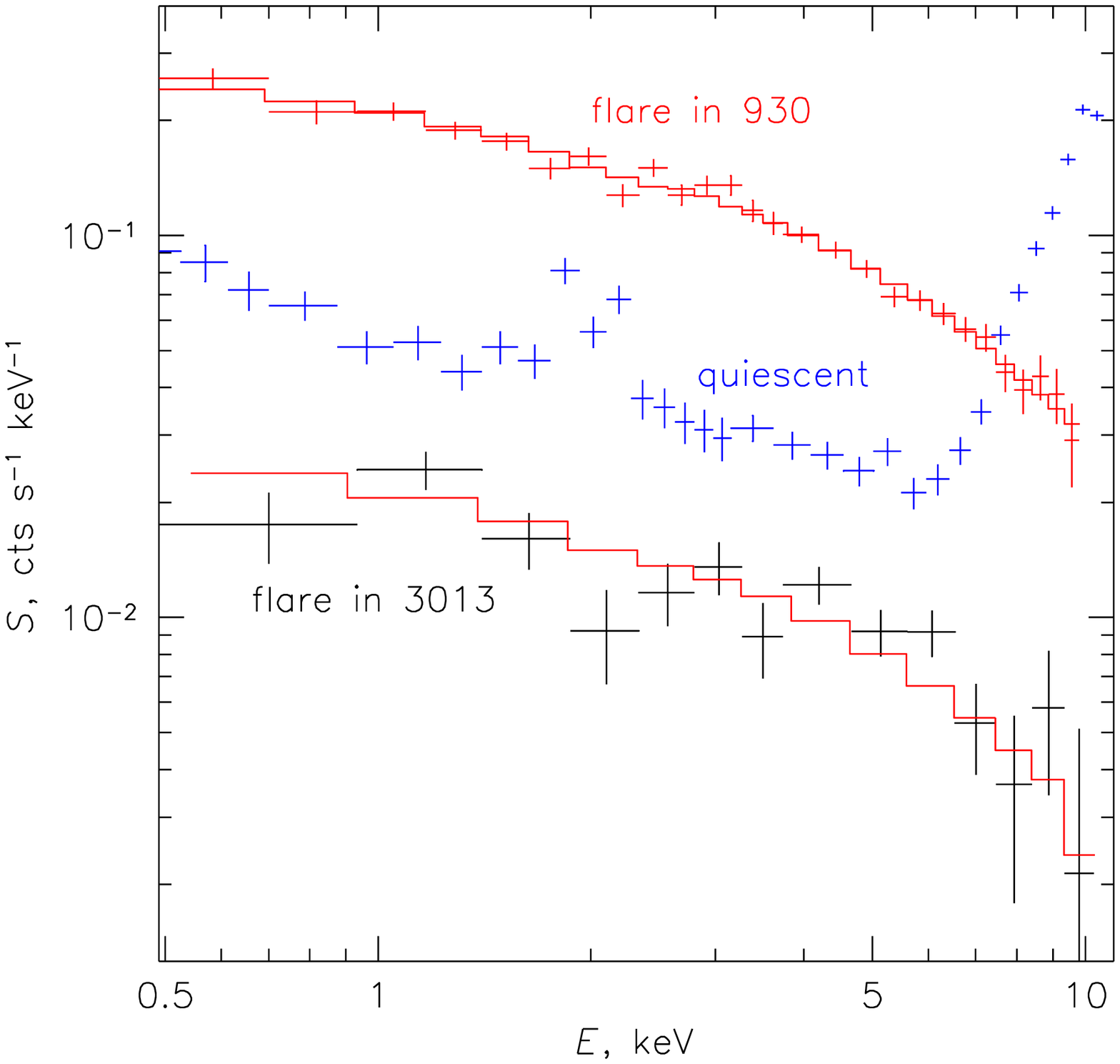}}

\rput[tl]{0}(0.0,15.4){
\begin{minipage}{8.75cm}
\small\parindent=3.5mm

\caption{Spectra of the excluded background flares (see text) in
  observations 930 (red) and 3013 (black), compared to the quiescent
  spectrum from the dark Moon (blue), for the whole S3 chip. Red histograms
  show a model that was fit to a combination of other observations with
  flares and renormalized, without a change of shape, to match the 930 and
  3013 spectra.}
\label{fig:flaresp}
\par
\end{minipage}
}
% \endpspicture
% \end{figure*}
% %%%%%%%%%%%%%%%%%%%%%%%%%%%%%%%%%%%%%%%%%%%%%%%%%%%%%%%%%%%%%%%%%%%%%%%%%%
% 
% 
% %%%%%%%%%%%%%%%%%%%%%%%%%%%%%%%%%%%%%%%%%%%%%%%%%%%%%%%%%%%%%%%%%%%%%%%%%%
% \begin{figure*}[t]
% \pspicture(0,20.2)(18.5,24.0)
% %\psgrid(0,0)(18.5,24)

\rput[tl]{0}(9.8,22){\epsfxsize=8.5cm \epsfclipon
\epsffile{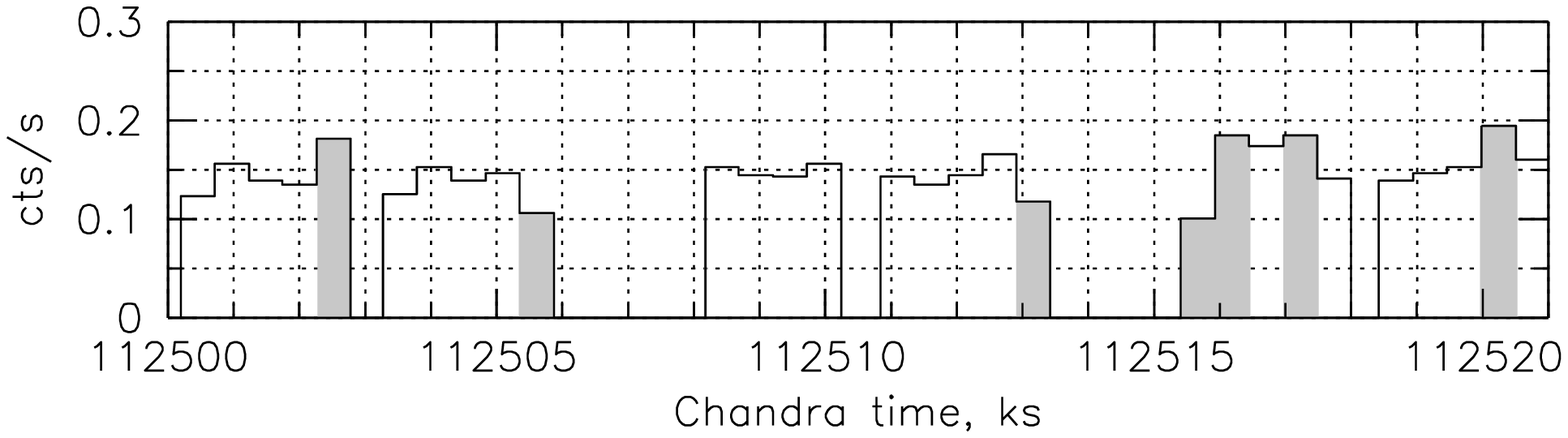}}

\rput[tl]{0}(9.75,19.2){
\begin{minipage}{8.75cm}
\small\parindent=3.5mm

\caption{The 2.5--7 keV light curve for the dark Moon observations.
  The rate is shown before the VF mode filtering (\S\ref{sec:vf}). Shaded
  bins are above or below a factor of 1.2 of the quiescent level and are
  excluded.}
\label{fig:lcmoon}
\par
\end{minipage}
}

\endpspicture
\end{figure*}

Again, we exclude the time periods above a factor of 1.2 of the apparent
quiescent rate. Assuming that the background excess in the rest of the
exposure is indeed a flare, we can try to model it by taking advantage of
the empirical finding, based on a number of observations, that the spectral
shape of the most frequent, ``soft'' species of the BI flares stays the same
even while their flux varies strongly in time. To illustrate this, we derive
spectra of the background flare components from the rejected, high count
rate periods of observations 3013 and 930. We first check the light curves
of FI chips also used in these observations, in order to exclude flares of a
different, ``hard'' species that is less frequent, affects both the BI and
FI chips and has a different spectrum. There are 1--2 ks of such flares
within the already excluded time periods in each observation. After
excluding those, we extract the spectra from the remaining high-rate
periods, and subtract from them the spectra of the quiescent periods of the
same observations, normalizing them by the respective exposure ratios
(thereby subtracting all time-independent sky and instrumental background
components). The resulting flare spectra are shown in
Fig.~\ref{fig:flaresp}.

In a similar manner, we also extracted flare spectra from several other
archival observations (OBSIDs 766, 326, 2206, 2076, 2213, and 1934) spanning
a period February 2000 -- September 2001 and a range of flare intensities.
Flares in all those observations can be described by a model consisting of a
power law with a photon index of $-0.15$ and an exponential cutoff at 5.6
keV, without the application of the telescope effective area and the CCD
quantum efficiency (commands `arf none' or `model/b' in XSPEC). In
Fig.~\ref{fig:flaresp}, this model is over-plotted on the flare spectra from
our CXB fields 930 and 3013, fixing the spectral shape and fitting only the
normalization.  As the figure shows, flares in both our observations have
nearly the same spectral shape, despite their time difference of 1.5 years
and very different intensities.  Freeing the spectral shape parameters, we
obtained a photon index of $-0.10\pm0.07$ and a cutoff at $5.2\pm0.7$ keV
for observation 930, and $-0.1\pm0.3$ and $7.2_{-3.4}^{+\infty}$ keV for
3013, consistent with the above fit for the composite flare spectrum.

Therefore, if the background excess affecting the useful period of
observation 3013 is indeed a residual flare, we can expect that it has the
same spectral shape, only a still lower normalization. As will be seen below
(\S\ref{sec:3013resid}), its spectrum is indeed consistent with this
assumption at the energies where the comparison is possible.  To try
modeling the residual background excess in 3013, we chose to fix the shape
parameters to the best-fit values from this particular observation (given
above), even though they are less strongly constrained than those from the
composite spectrum, to account for any possible slow evolution of the
spectrum which cannot be ruled out with the data at hand. Of course this
choice has no significant effect on our results.  The normalization of this
flare model will be determined below (\S\ref{sec:3013resid}) after
subtracting the quiescent background component and removing point sources.

For comparison, Figure~\ref{fig:flaresp} also shows a quiescent instrumental
background spectrum from the dark Moon observations (discussed below).  We
note that the BI ``soft'' flare spectrum at high energies is much softer
than the quiescent background; unless the flare is very strong, its
contribution above 10 keV is unnoticeable (this is not true for the other,
``hard'' flare species mentioned above). This fact will be used below.

\subsection{Quiescent background}

\subsubsection{Dark Moon observations}
\label{sec:moon}

To separate the CXB and instrumental components of the ACIS background,
\chandra\ observed the dark Moon in July 2001 in a series of 6 short
pointings (OBSIDs 2469, 2487, 2488, 2489, 2490, 2493), tracking the Moon for
a total of about 15 ks. ACIS chips S2, S3, I2 and I3 were on and telemetered
data in Very Faint (VF) mode. A second installment of Moon observations in
September 2001 exposed chips I2 and I3. Technical difficulties encountered
in these two runs, related mostly to the fact that \chandra\ aspect camera
cannot be used near the Moon, prevented further dark Moon observations. Here
we use only the S3 data from July 2001.  During that run, optical flux from
the 1/3 of the Lunar disk that was illuminated was imaged onto to the ACIS
focal plane. The ACIS optical blocking filters were not designed to reject
visible light from the sunlit Moon, and a detectable offset signal was
produced. The effect was most severe in chip I2, but very small in chip S3.
A correction to each event's Pulse Height Amplitude (PHA) was calculated
individually by averaging the lowest 16 pixels of the $5\times 5$ pixel VF
mode event island.  The bias error in S3 was well below the threshold of
affecting the event grades, so this problem did not result in any loss of
events due to the onboard grade rejection (as was the problem for chip I2 in
this dataset). The average correction to the energy for events in chip S3
was within a few eV, negligible for our purposes.

A 2.5--7 keV S3 light curve for all July 2001 dark Moon observations is
shown in Fig.~\ref{fig:lcmoon}. The end of the exposure was affected by an
apparent faint flare which was filtered using the same factor of 1.2
threshold as in \S\ref{sec:lc} (we had to use smaller time bins which
resulted in some statistical deviations that were also excluded for
consistency). The resulting clean Moon exposure for S3 is 11400 s. At high
energies where the CXB contribution is negligible, the Moon quiescent
background rate was within a few percent of that in other recent
observations. Scientific results from the Moon observations will be
discussed by C. E. Grant et al.\ (in preparation).

\subsubsection{Event Histogram Mode data}

An independent approach to calibrate the ACIS instrumental background
utilizes the Event Histogram Mode (EHM) data (Biller et al.\ 2002). These
data are collected during science observations by the HRC-I detector while
ACIS is stowed inside the detector support structure.  This structure blocks
celestial X-rays but does not affect the particle rate significantly (as
will be seen below). In this mode, the telemetry capacity available for ACIS
is small, so the only information transmitted is a PHA histogram for the
events from a predefined region of the chip. For this reason, the usual
exclusion of bad pixels and the position-dependent gain correction cannot be
applied. At this location, ACIS is also faintly illuminated by the internal
calibration line source. The flare component discussed in \S\ref{sec:lc} is
never observed in the stowed position.

\begin{figure*}[t]
\pspicture(0,14.2)(18.5,24.0)
%\psgrid(0,0)(18.5,24)

\rput[tl]{0}(1.,24){\epsfxsize=16cm \epsfclipon
\epsffile[18 204 588 500]{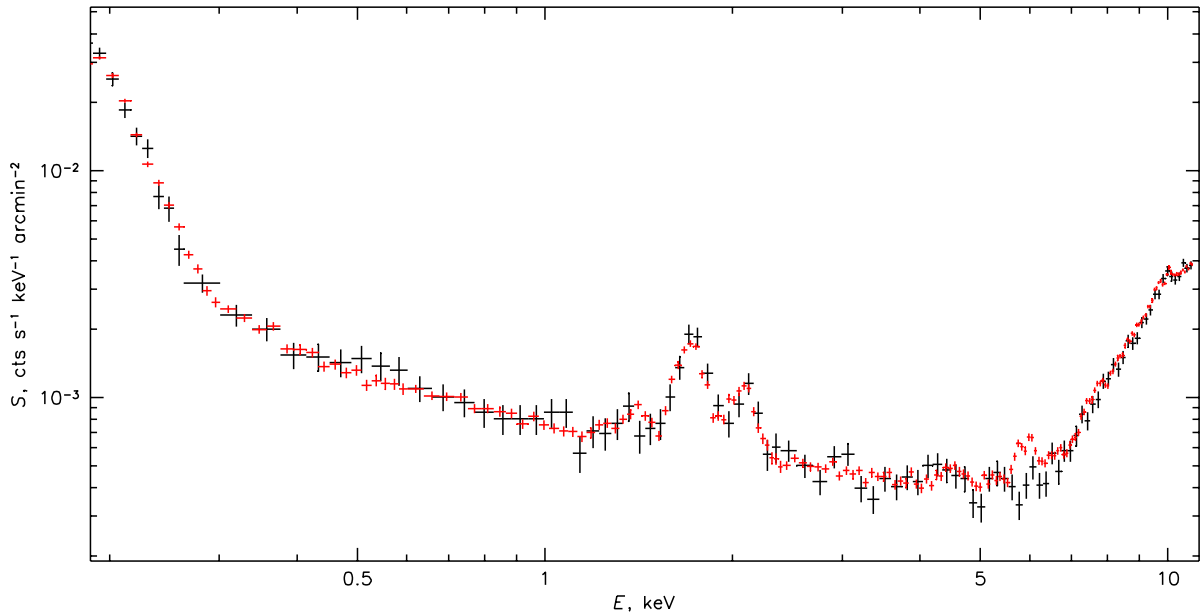}}

\rput[tl]{0}(0.0,15.4){
\begin{minipage}{18.5cm}
\small\parindent=3.5mm

\caption{Spectra of dark Moon (black) and Event Histogram Mode data (red).
  This Moon spectrum is extracted in PHA channels in a special manner to be
  directly comparable to the EHM data (see text).  The energy scale is
  approximate (PHA values multiplied by 4.7 eV per channel).  Apart from the
  faint internal calibration source lines present in the EHM data (most
  notably at 5.9 keV and 1.5 keV), there is very good agreement between the
  spectra.}
\label{fig:moonhist}
\par
\end{minipage}
}

\endpspicture
\end{figure*}

An EHM spectrum from the whole S3 chip accumulated over the July-October
2001 period (straddling the date of the Moon observation) is shown in
Fig.~\ref{fig:moonhist} (see also Biller et al).  For comparison, we
overlaid a dark Moon spectrum. In order for it to be directly comparable to
the EHM spectrum, in deriving it we did not exclude bad CCD pixels, apply
gain corrections (that is, we used PHA rather than Pulseheight Invariant, or
PI, values), or apply the additional VF mode filtering (\S\ref{sec:vf}).

The figure shows that away from the calibration source lines, the agreement
between the spectra is quite remarkable over the entire energy range, within
the statistical accuracy of the Moon dataset.  One might expect both the
dark Moon and the EHM spectra to exhibit emission above the non-cosmic
background level seen in ordinary observations --- e.g., \rosat\ and \asca\
Moon data suggested emission at low energies (Schmitt et al.\ 1991; Kamata
et al.\ 1999), and the detector support structure may be radioactive. In
principle, one could also imagine a component of the quiescent particle
background that can be blocked by that structure.  While the coincidence of
the two spectra in Fig.~\ref{fig:moonhist} does not rule out a conspiracy of
these possibilities, it makes each of them very unlikely. It also supports
the conclusion by Freyberg (1998) that the \rosat\ emission in the direction
of the dark Moon was actually fluorescent emission from a region around the
Earth (i.e., below the \chandra\ orbit).  Therefore, we will assume that
both the dark Moon and the EHM data give the true quiescent background, and
use the Moon dataset as an instrumental background model for the sky data
below.

The EHM data were also used directly as the background model for a CXB study
by Edgar et al.\ (2002).  Another measure of the ACIS background was
obtained in August 1999 just prior to opening the \chandra\ telescope
door. That dataset is analyzed in Baganoff (1999) which may be consulted for
background line identifications and other qualitative information.
Unfortunately, those data cannot be used directly for recent observations,
because both the background and the ACIS detector have evolved significantly
since then. Calibration observations with ACIS stowed and working in the
full imaging mode (as opposed to EHM), forthcoming in 2002 and 2003, could
be used for the most recent observations.

\subsubsection{Background time dependence}
\label{sec:timedep}

\begin{figure*}[t]
\pspicture(0,14.0)(18.5,24.0)
%\psgrid(0,13)(18.5,24)

\rput[tl]{0}(-0.2,24){\epsfxsize=9.0cm\epsfclipon
\epsffile{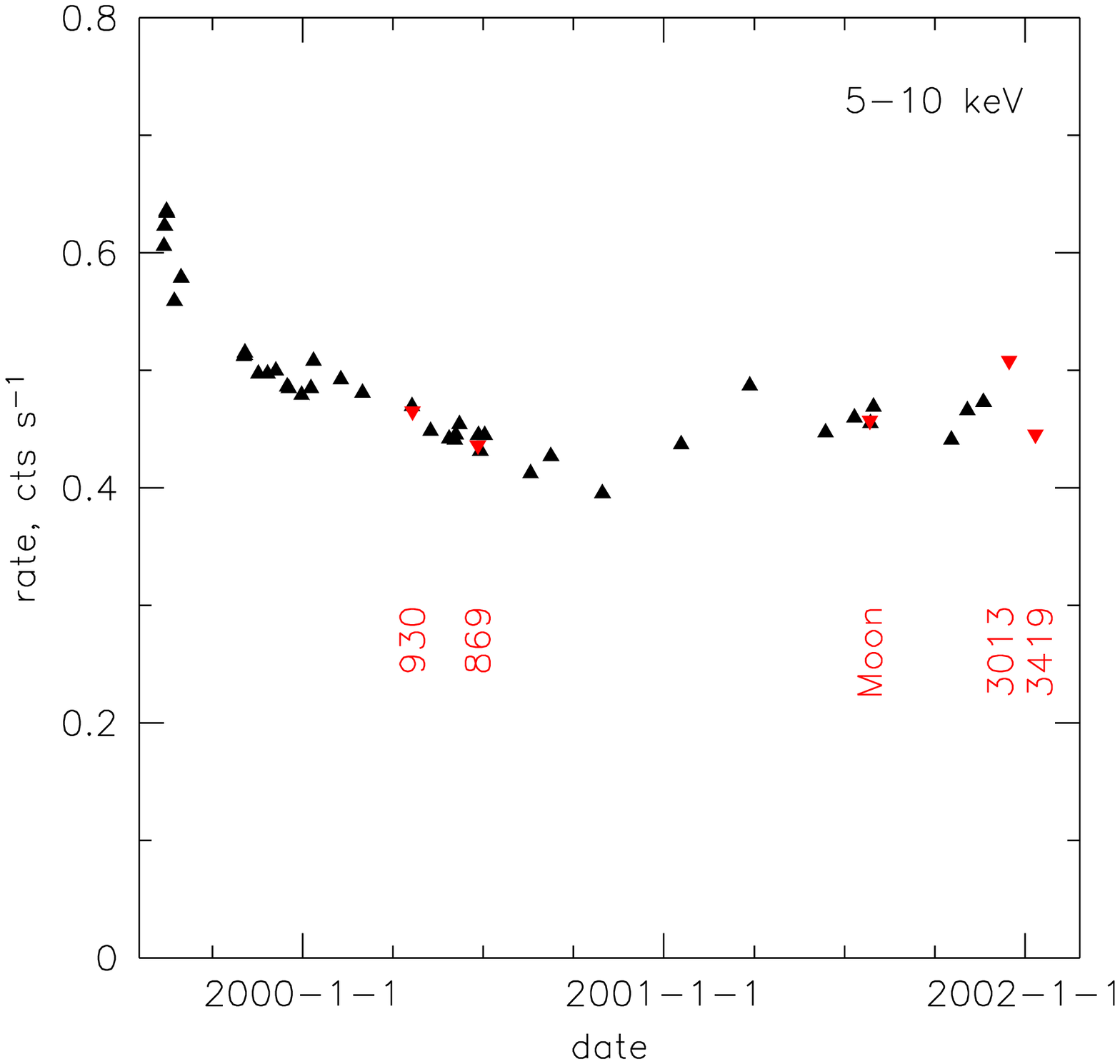}}

\rput[tl]{0}(0.0,15.4){
\begin{minipage}{8.75cm}
\small\parindent=3.5mm

\caption{Time dependence of the S3 chip 5--10 keV quiescent background rate
  in various blank field observations since launch (see Markevitch
  2001). Statistical errors are comparable to the symbol size. Red symbols
  denote observations used in this work (OBSID 3013 is shown before the
  residual flare correction).}
\label{fig:timedep}
\par
\end{minipage}
}

%%%%%%%%%%%%%%%%%%%%%%%%
\rput[tl]{0}(9.8,24){\epsfxsize=9cm \epsfclipon
\epsffile{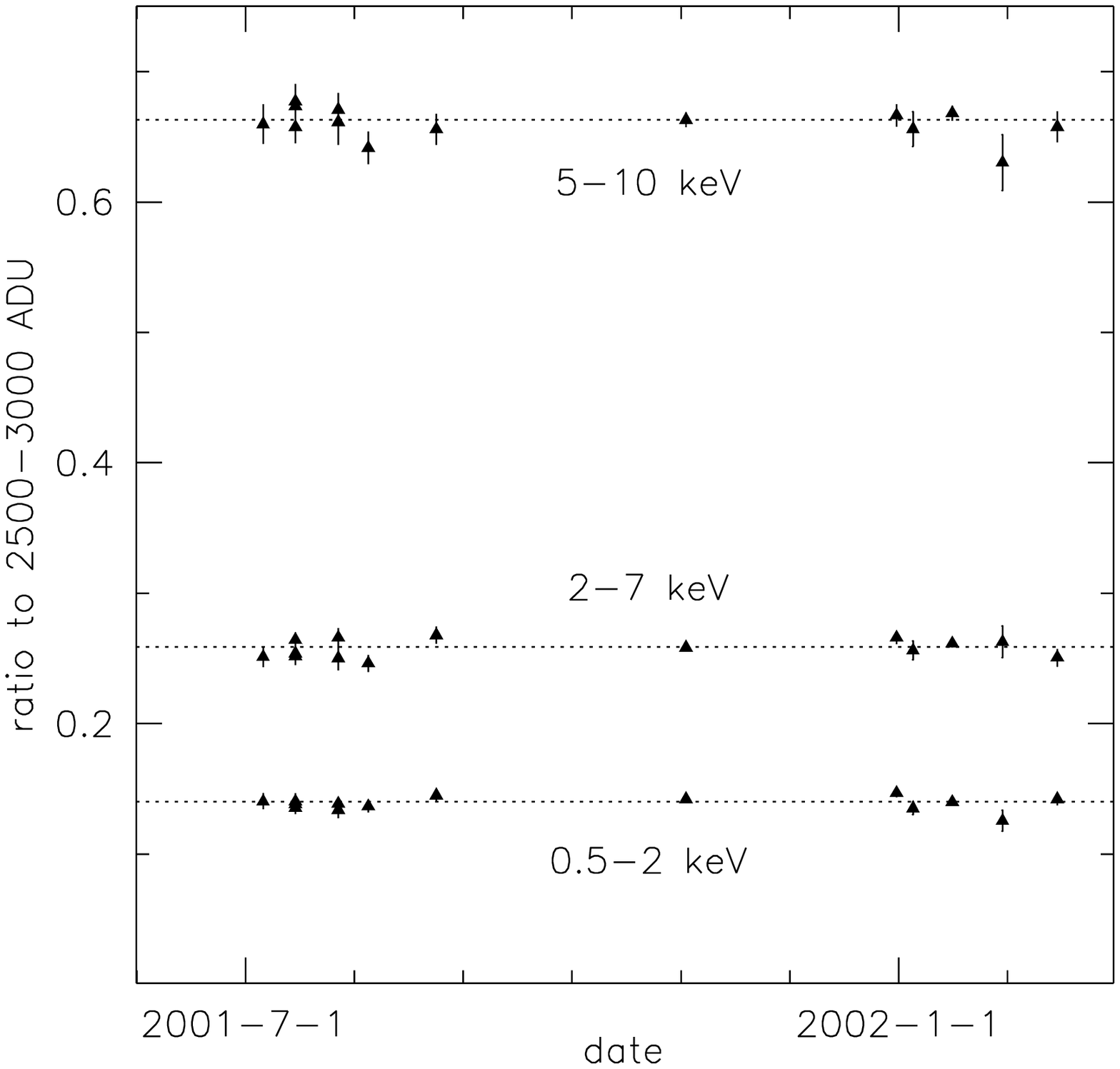}}

\rput[tl]{0}(9.75,15.4){
\begin{minipage}{8.75cm}
\small\parindent=3.5mm

\caption{Ratios of the S3 background rates in the 0.5--2 keV, 2--7 keV, and
  5--10 keV bands to the 2500--3000 ADU ($\sim$ 10--12 keV) rate, for
  EHM observations that used the full chip. Horizontal lines show
  average values.}
\label{fig:correl}
\par
\end{minipage}
}
\endpspicture
\end{figure*}

From the analysis of a large number of ACIS observations and monitoring of
the rate of ACIS events rejected onboard (Grant, Bautz, \& Virani 2002), we
know that the high-energy quiescent background slowly declined since launch
until around the end of 2000, and has been relatively constant during 2001
(Figure~\ref{fig:timedep} shows this behavior in the 5--10 keV band where
the instrumental background dominates), in anticorrelation with the solar
cycle.  At low energies, the qualitative behavior is similar, but it is
difficult to tell exactly because of the differing sky signal. In addition
to the slow evolution, there are small variations of the quiescent rate on
short time scales. Since the background may change between the observations
of our fields and the Moon, the Moon background may need a correction, and
the background uncertainty must be included in the final results.

The EHM data are free from the sky signal and flares and may be used for
checking the quiescent background time dependence, especially in the soft
band most important for the present study.  From the comparison of EHM (as
well as blank field) data in different energy bands, it appears that during
the short-term quiescent background fluctuations such as those in
Fig.~\ref{fig:timedep}, its spectral shape does not change significantly and
only the normalization varies. This behavior enables us to account for these
variations by normalizing the model instrumental background by the ratio of
rates at high energies, e.g., 10--12 keV, where the contribution from
celestial sources and possible faint undetected flares (\S\ref{sec:lc3013})
is negligible.

Figure~\ref{fig:correl} shows EHM rates from the full S3 chip in the 0.5--2
keV, 2--7 keV, and 5--10 keV bands divided by the rates in the PHA interval
of 2500--3000 Analog-Digital Units, or ADU (approximately 10--12 keV; PHA is
preferred over energy because many science observations use an onboard
cutoff at 3000 ADU). The period from July 2001 is shown because earlier EHM
data were collected from parts of the chip. The scatter of these ratios is
quite small; in all energy bands, the intrinsic scatter around the mean
required in addition to the Poissonian scatter is 1.5--2\% ($1\sigma$) or
less, a reduction from 3--4\% for fluxes not normalized by the high-energy
rate. One also notices that neither the rates nor the spectral shape changed
systematically between the dates of the Moon observation and our deep
observations 3013 and 3419.

We use such high-energy rate matching in our CXB analysis by normalizing the
Moon spectrum by the ratio of the respective 2500--3000 ADU rates. For
OBSIDs 3013, 3419, 930, and 869, such normalizations are 1.06, 0.97, 1.01,
and 0.93 times the ratio of the Moon exposure to the respective exposures,
so this correction is, as expected, small. We adopt a systematic uncertainty
of the resulting quiescent background normalization of 2\% ($1\sigma$) as
derived above.

We note that over longer periods, the spectral shape of the detector
background may change --- the spectra of the empty field observations from
2000 and 2001 differ by $\sim 5$\% at energies where the non-cosmic
component dominates. This means that the Moon spectrum with a simple
normalization adjustment may not be a good model for our earlier
observations 930 and 869. However, as will be seen below, the soft diffuse
signal in those observations is so strong that the background uncertainty
does not matter.

\subsection{VF mode background filtering}
\label{sec:vf}

\begin{figure*}[t]
\pspicture(0,14.2)(18.5,24.0)
%\psgrid(0,13)(18.5,24)

\rput[tl]{0}(0.,24){\epsfxsize=9.0cm \epsfclipon
\epsffile{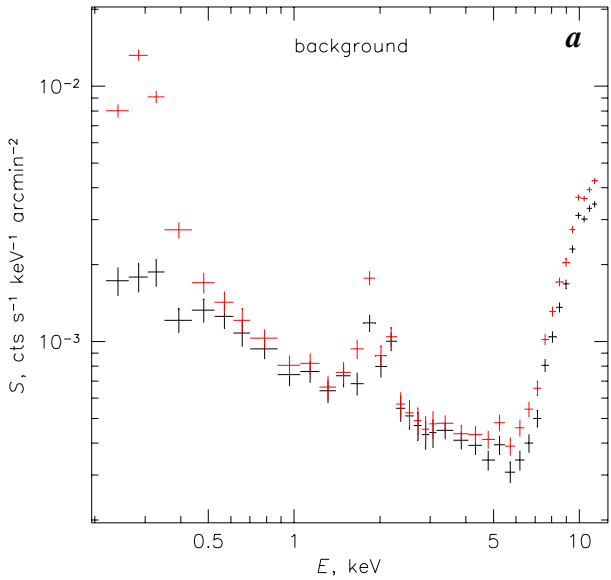}}

\rput[tl]{0}(9.5,24){\epsfxsize=9.0cm \epsfclipon
\epsffile{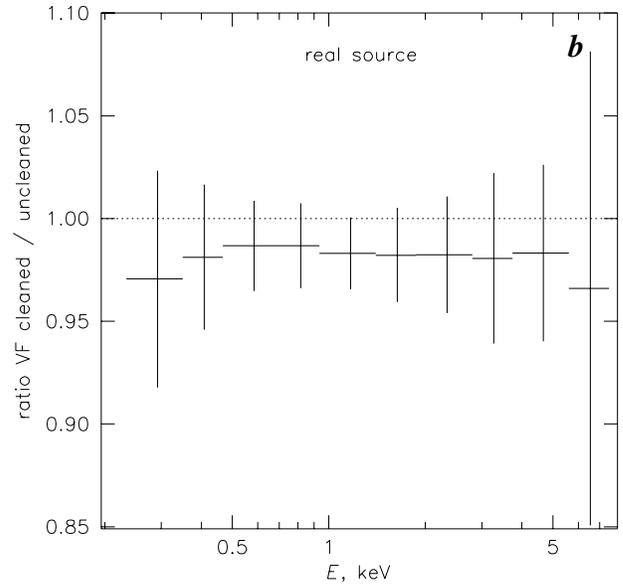}}

\rput[ll]{0}(7.7,22.8){\large\bi a}
\rput[ll]{0}(17.1,22.8){\large\bi b}

\rput[tl]{0}(0.0,15.4){
\begin{minipage}{18.5cm}
\small\parindent=3.5mm

\caption{({\em a}) Spectra of the S3 detector background (dark Moon)
  before (red) and after (black) VF mode cleaning. ({\em b}) Ratio of the
  spectra of a bright celestial source (without pileup) after and before VF
  mode cleaning. The non-cosmic background is significantly reduced at low
  and high energies, while the effect on the real X-rays is
  energy-independent and very small.}
\label{fig:vf}
\par
\end{minipage}
}
\endpspicture
\end{figure*}

In Very Faint ACIS telemetry mode, the detector background can be reduced
significantly by rejecting events with signal above the split threshold in
any of the outer pixels of the $5\times 5$ pixel event island, after an
approximate correction for the charge transfer inefficiency. Details of the
method can be found in Vikhlinin (2001).  Figure~\ref{fig:vf} shows the
effect of such filtering on the spectra of the background (the dark Moon)
and real X-ray events from an extended celestial source unaffected by photon
pileup. It results in a significant reduction of the background rate,
especially at the lowest and highest energies, while rejecting only about
2\% of the real events. The background reduction is stronger for FI chips
(not used in this work). All four of our CXB observations, as well as all
Moon observations, were telemetered in VF mode and filtered in this manner.

\section{POINT SOURCES}
\label{sec:src}

We will now detect and exclude point sources in our CXB fields.  All of the
S3 chip is within $7'$ of the optical axis and the PSF is narrow over the
whole field, so point source detection is photon-limited and background is
relatively unimportant for it. It is therefore advantageous to include
periods of moderately high background if it significantly increases the
exposure. For observations 3013, 3419 and 930, we applied a less restrictive
light curve filtering (using the 0.3--10 keV band, and for 930, a higher
threshold factor of 2 instead of 1.2) to increase the exposures for source
detection (see Table~\ref{table:obslist}).

In the 3 fields with point-like original targets (OBSIDs 3013, 3419, 930),
we excluded $r=30''$ circles around the target from the analysis. In OBSID
869, whose original target is a galaxy, we masked that source liberally and
used the remaining 72\% of the chip area.

Source detection was performed in two bands, 0.3--2 keV and 2--7 keV. The
source candidates were identified in a standard manner by applying wavelet
filtering to the image to reduce statistical noise and then searching for
local brightness maxima using the code of Vikhlinin et al.\ (1998). The
source positions were then refined by photon centroiding, and the source
fluxes calculated within the 90\% PSF encircled energy radii $r_{90}$ (and
then divided by 0.9) using as the local background a wavelet decomposition
component containing details on the largest linear scale. For simplicity, we
set a relatively high, spatially uniform lower limit on the source flux that
corresponds approximately to 8--10 photons anywhere in the field in each
observation. This ensures that all detected sources are real and the
background contribution to the source flux is always small, even though we
may miss some obvious fainter sources near the optical axis. As will be seen
below, these omissions do not affect our results significantly.

\begin{figure*}[t]
\pspicture(0,13.5)(18.5,24.0)
%\psgrid(0,0)(18.5,24)

\rput[tl]{0}(-0.1,24.5){\epsfxsize=9.0cm \epsfclipon
\epsffile{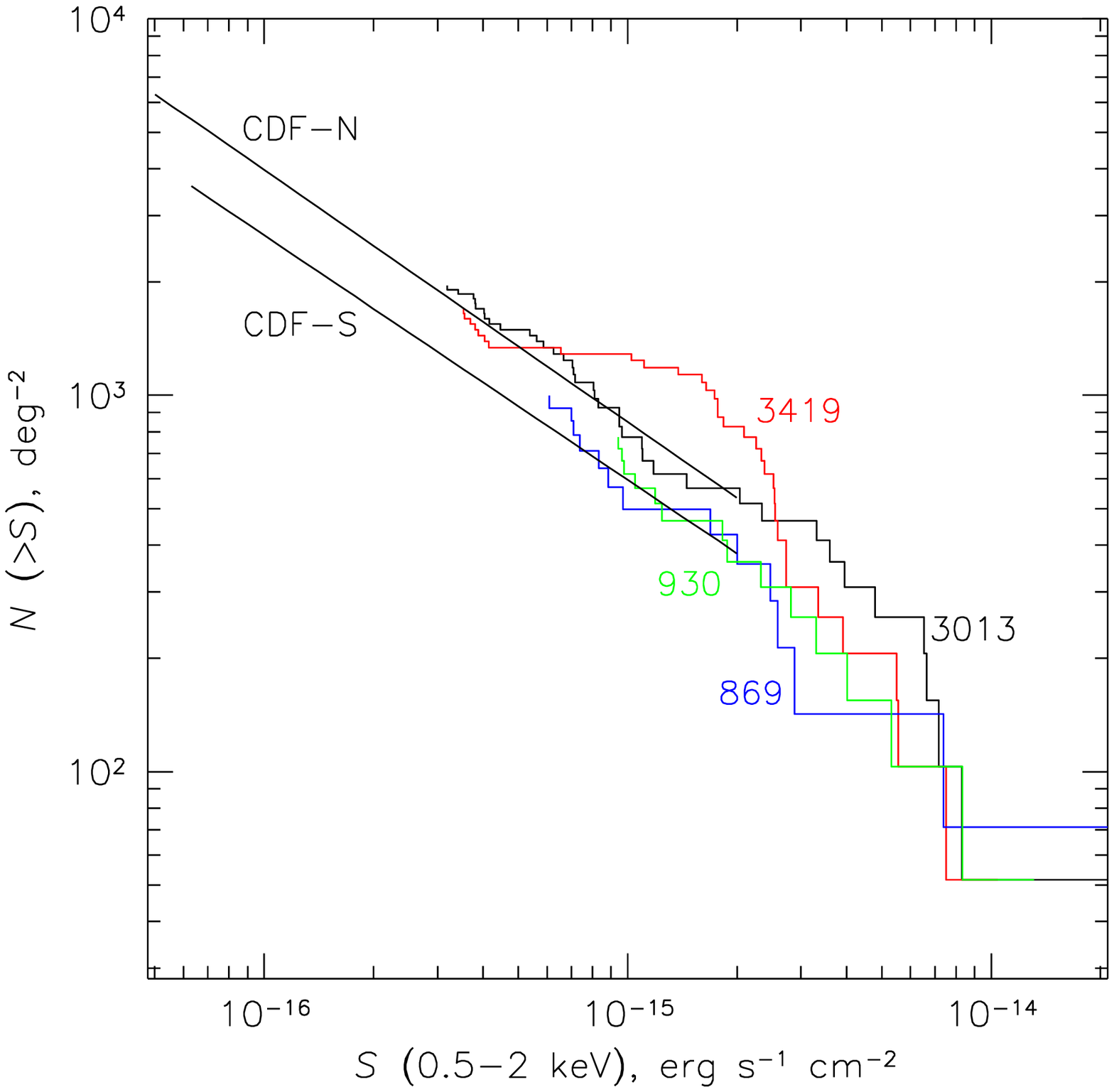}}

\rput[tl]{0}(0.0,15.8){
\begin{minipage}{8.75cm}
\small\parindent=3.5mm

\caption{Cumulative numbers of excluded point sources as a function
  of their 0.5--2 keV flux. Only the sources detected in the 0.3--2 keV band
  are shown. Labels give OBSIDs. The curves end at our adopted lower flux
  cuts, which approximately correspond to 8--10 photons from the source. For
  comparison, low-flux fits to the CDF-N (Brandt et al.\ 2001) and CDF-S
  (Rosati et al.\ 2002) source counts are shown.}
\label{fig:counts}
\par
\end{minipage}
}
%%%%%%%%%%%%%%%%%%%%%%%%%%%%%%%%%%%%
\rput[tl]{0}(9.8,24.5){\epsfxsize=9.0cm \epsfclipon
\epsffile{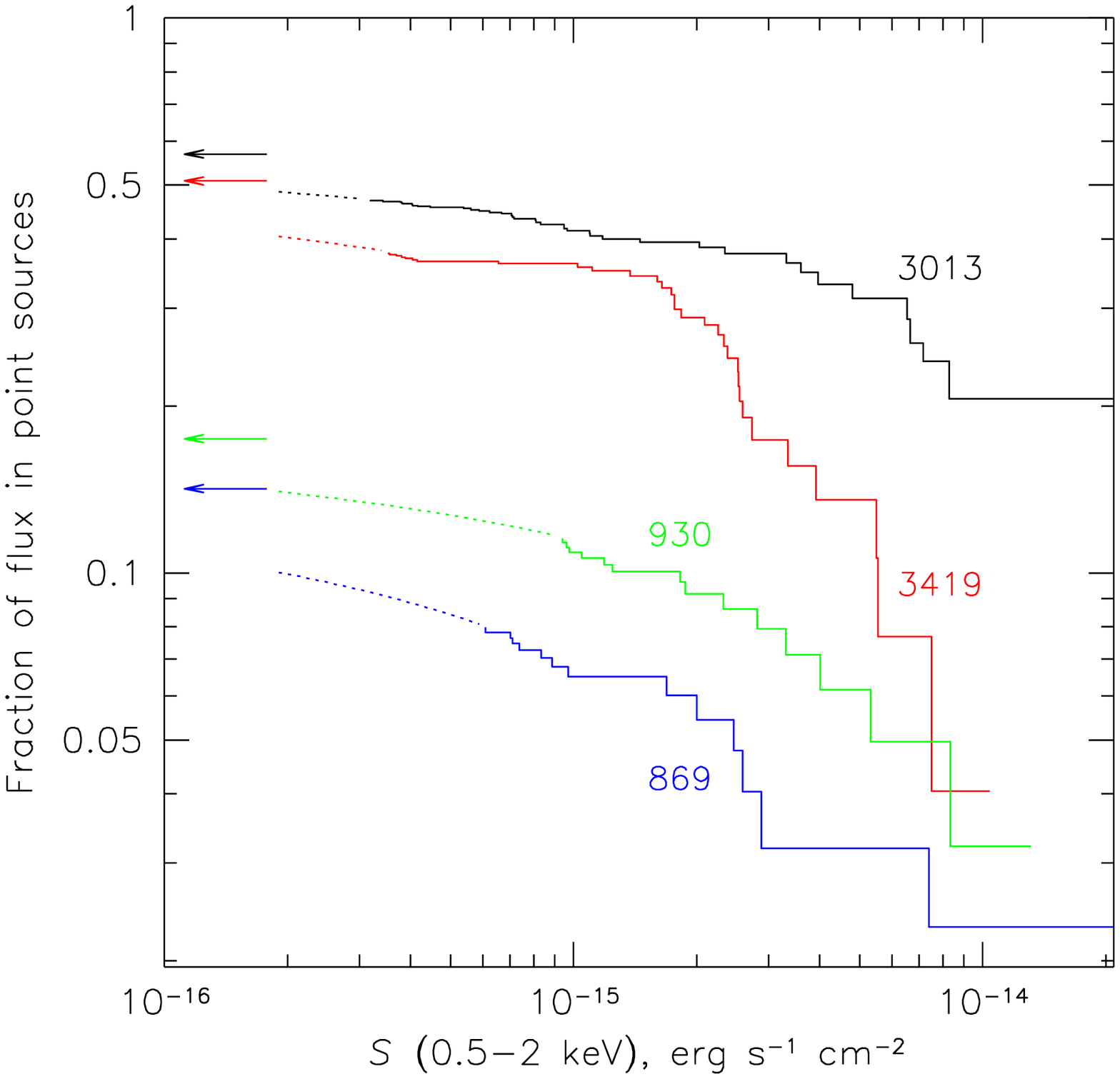}}

\rput[tl]{0}(9.75,15.8){
\begin{minipage}{8.75cm}
\small\parindent=3.5mm

\caption{Contribution of the detected point sources above a
  certain flux to the total 0.3--2 keV count rate (minus detector
  background), as a function of the source 0.5--2 keV flux. Labels give
  OBSIDs. Dotted lines show extrapolations toward lower fluxes, assuming
  source counts as in CDF-N (Fig.~\ref{fig:counts}); arrows indicate their
  asymptotic limits at zero flux.}
\label{fig:cumflux}
\par
\end{minipage}
}

\endpspicture
\end{figure*}

The resulting cumulative source counts from the 0.3--2 keV images as a
function of the unabsorbed 0.5--2 keV source flux (assuming Galactic
absorption and a power law with a $-1.4$ photon index for the source
spectrum, relevant for the faintest sources) are presented in
Fig.~\ref{fig:counts}. The figure also shows fits to the low-flux end of the
0.5--2 keV source counts from much deeper observations of \chandra\ Deep
Field North (CDF-N; Brandt et al.\ 2001) and South (CDF-S; Rosati et al.\
2002) who used the same assumption about the average source spectrum.  Our
curves are in good agreement with those results.  The field-to-field
difference (a factor of $\sim 2$) is also similar to that between CDF-N and
CDF-S (see also Barcons, Mateos, \& Ceballos 2000), although with such low
absolute source numbers this result is not particularly significant.

For the interpretation of our diffuse measurements, it is useful to estimate
the expected contribution of point sources below our detection limits.
Figure~\ref{fig:cumflux} shows the cumulative 0.3--2 keV flux of all sources
above a certain flux, divided by the total CXB signal in each observation
(the total flux minus the Moon background; this quantity is different in all
observations, being higher for OBSIDs 869 and 930). We can extrapolate these
curves below our limits assuming, for example, the faint source distribution
from CDF-N, as shown in the figure.  The arrows show the asymptotic limits
of those extrapolations at the zero flux. One can see that at most 60\% of
the 0.3--2 keV total CXB flux in our two main observations can be due to
point sources (barring the emergence of an unknown source population at
fluxes below the CDF limits, or a significant change in the average source
spectrum at the lowest fluxes; such possibilities are beyond the scope of
this paper).  The contribution of sources below our detection limits is less
than 10\% of the total CXB flux (and less than 20\% of the unresolved flux)
in all four observations, thus the uncertainty of our extrapolation toward
lower fluxes should not affect any of our conclusions.

At energies above 2 keV, the accuracy of our present measurements does not
warrant a detailed analysis of the point source contribution.  Lists of
sources detected in the 0.3--2 keV and 2--7 keV bands were merged, and
circles of radius $2r_{90}$ around the sources ($3r_{90}$ for sources with
$>100$ counts) were excluded from the spectral analysis below. Table
\ref{table:obslist} gives the resulting solid angles after the target and
source exclusion.

\section{SPECTRA AND INSTRUMENT RESPONSES}
\label{sec:rmf}

Diffuse CXB spectra were extracted in PI channels from the whole S3 chip
excluding the source regions. To extract the instrumental background spectra
from exactly the same chip areas, we converted the dark Moon event list into
the corresponding sky coordinate frame using the aspect track of each CXB
observation, assigning each Moon event a time tag selected randomly from the
time interval spanned by that observation. The dark Moon spectrum was then
extracted from the same region specified in sky coordinates, and normalized
as described in \S\ref{sec:timedep}.  All observations, including of the
Moon, are performed at the same focal plane temperature ($-120^\circ$C), so
the Moon spectra have the same resolution and can be directly subtracted
from the CXB spectra.

A spectral redistribution matrix (RMF) for each observation was calculated
by averaging the position-dependent matrices over the extraction region. 
Auxiliary response files (ARFs) included the telescope effective area and
CCD quantum efficiency averaged over the extraction region. Regions of the
masked sources were also excluded from the response averaging (this has any
effect only for OBSID 869 where a relatively large area is excluded).  ARFs
and RMFs were calculated using A. Vikhlinin's tools `calcarf' and `calcrmf'.

A recently discovered slow systematic decline of the ACIS quantum efficiency
at low energies (Plucinsky et al.\ 2002) was taken into account in the ARFs.
The present calibration accuracy of the quantum efficiency at $E\simeq
0.5-0.7$ keV is $\sim 10$\%.

\section{RESIDUAL BACKGROUND EXCESS}
\label{sec:3013resid}

\begin{figure*}[t]
\pspicture(0,5.4)(18.5,23.4)
%\psgrid(0,0)(18.5,24)

\rput[tl]{0}(0.5,24){\epsfxsize=9.0cm \epsfclipon
\epsffile{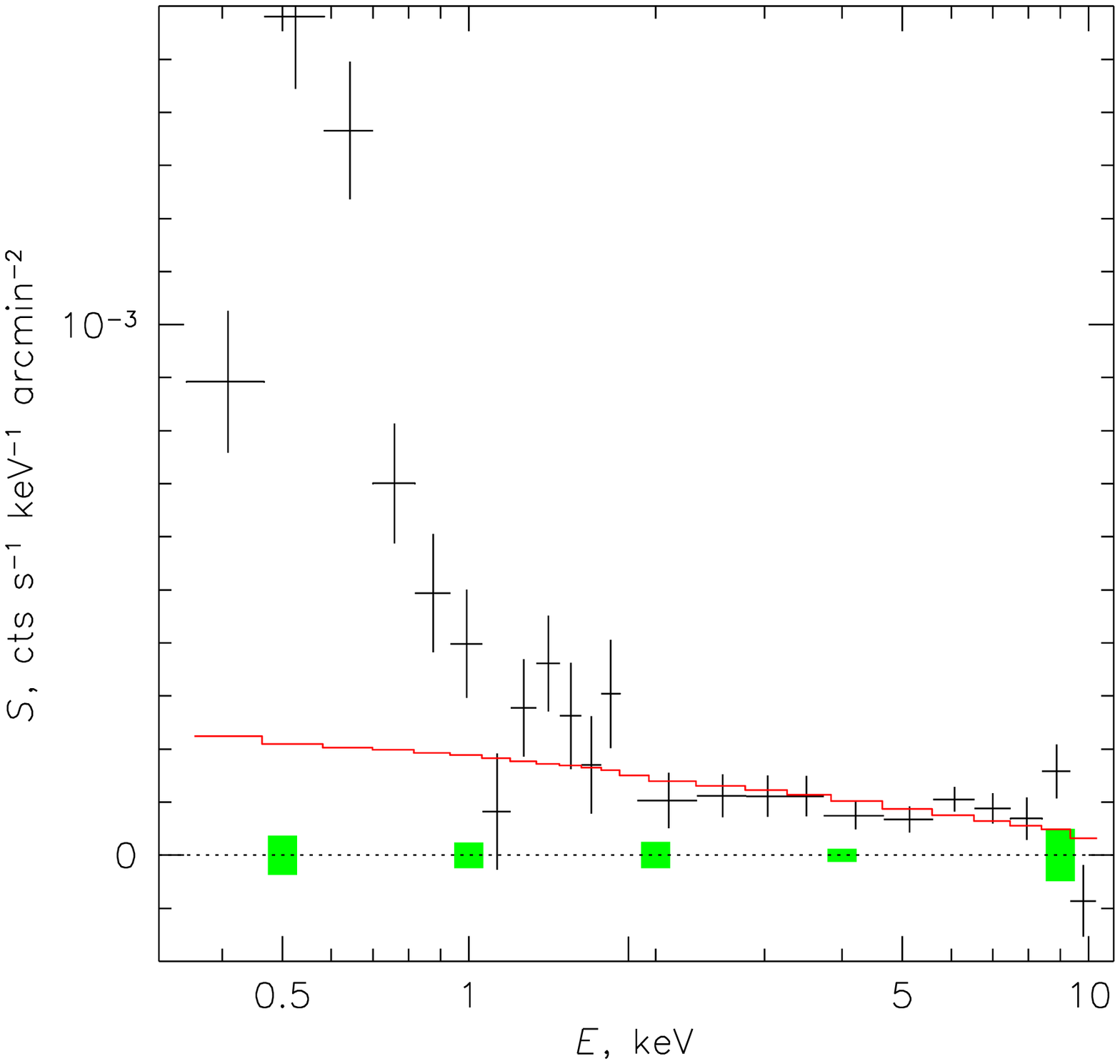}}

\rput[tl]{0}(9.5,24){\epsfxsize=9.0cm \epsfclipon
\epsffile{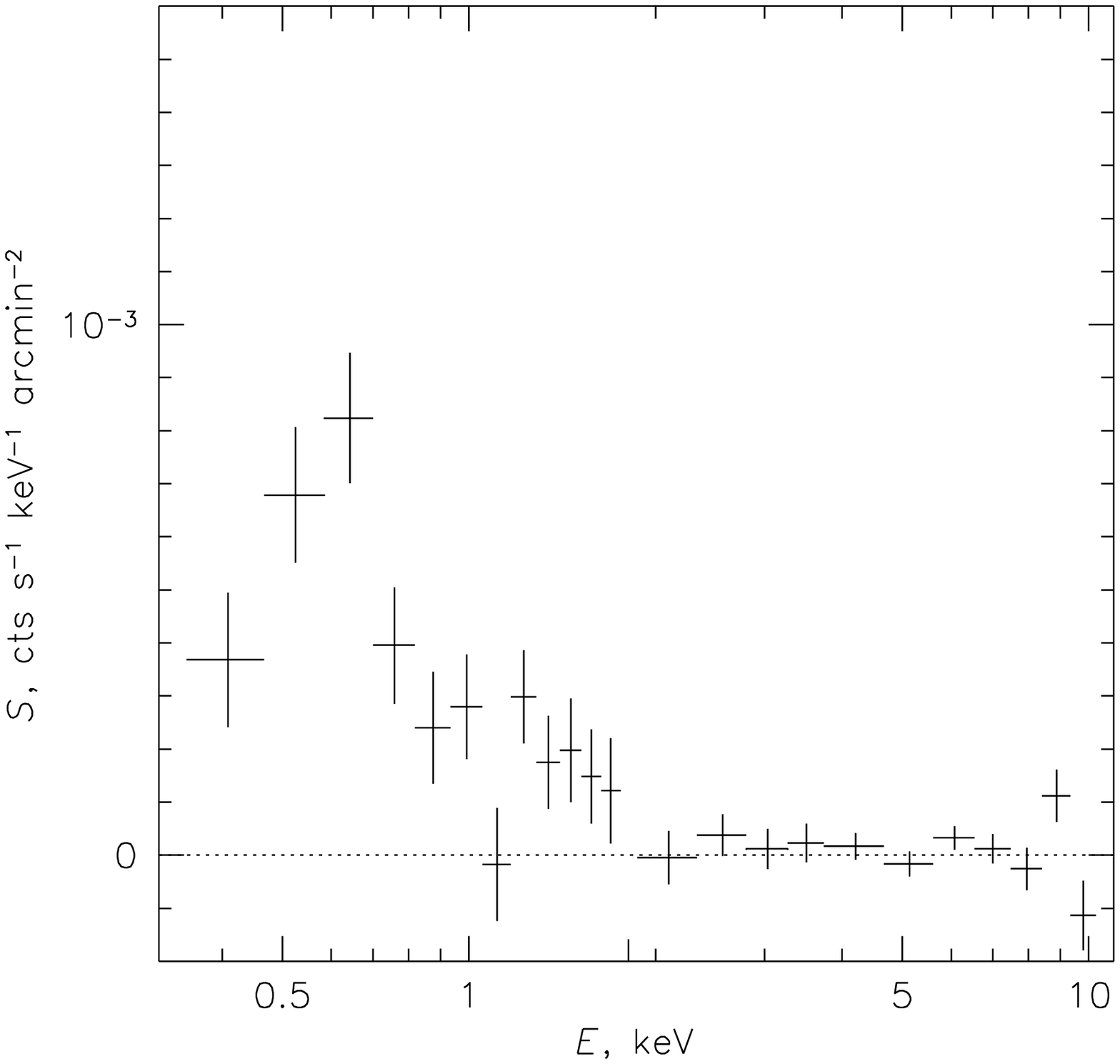}}

\rput[tl]{0}(0.5,15.8){\epsfxsize=9.0cm \epsfclipon
\epsffile{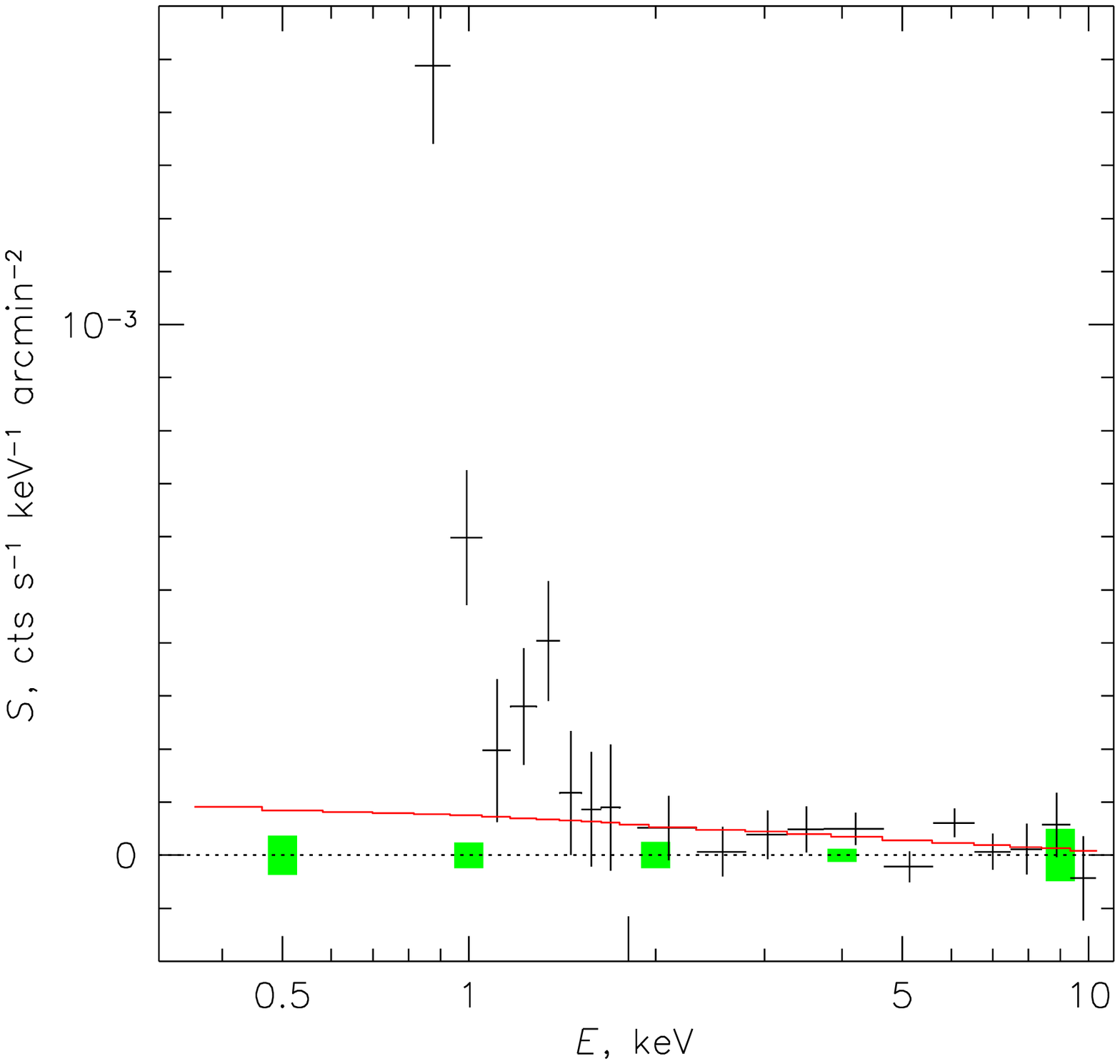}}

\rput[tl]{0}(9.5,15.8){\epsfxsize=9.0cm \epsfclipon
\epsffile{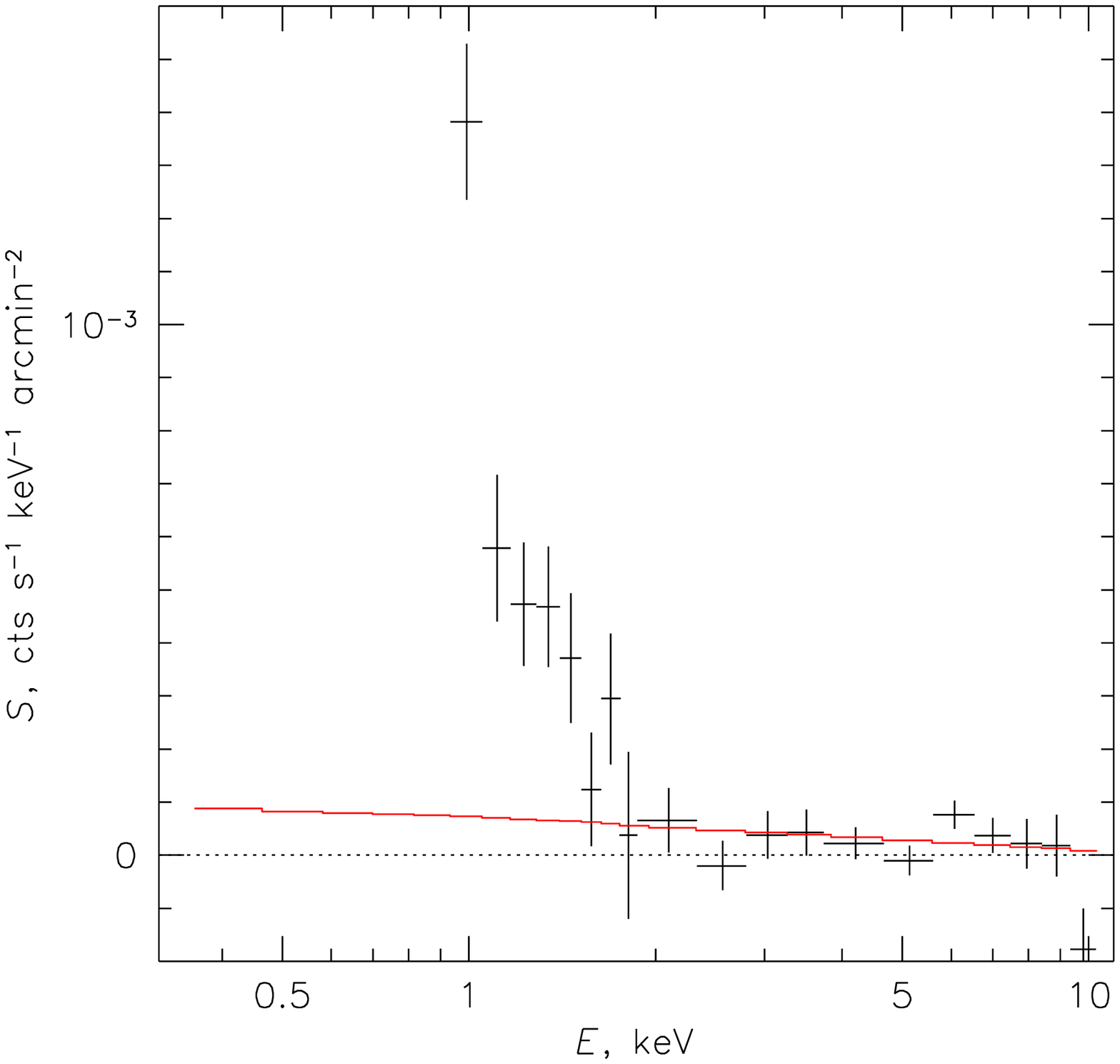}}

\rput[br]{0}(8.4,22.6){\bf 3013}
\rput[br]{0}(17.4,22.6){\bf 3419}
\rput[br]{0}(8.4,14.4){\bf 869}
\rput[br]{0}(17.4,14.4){\bf 930}

\rput[tl]{0}(0.0,7.0){
\begin{minipage}{18.5cm}
\small\parindent=3.5mm

\caption{First-iteration spectra of the diffuse component for the four
  fields. OBSIDs are marked in each panel. For OBSIDs 3013, 869 and 930,
  models of the residual flares, whose normalizations were fit in the
  2.5--10 keV band, are shown as red histograms (see text). The soft flux in
  fields 869 and 930 is above the limit of the plots. Green bars illustrate
  the 90\% quiescent background normalization uncertainty of $\pm 3$\%. In
  3013, the flare component is obvious and well described by the model; it
  is subtracted as an additional background component. In 869 and 930, it is
  marginally consistent with 0, given the background uncertainty. In all
  observations, the (possible) flare components are small compared to the
  diffuse signal in the soft band.}
\label{fig:residflare}
\par
\end{minipage}
}

\endpspicture
\end{figure*}

Our first-iteration spectra of the unresolved CXB from each observation are
shown in Fig.~\ref{fig:residflare}. For illustration, a 90\% systematic
uncertainty of the quiescent background normalization of $\pm3$\% is also
shown.  There is one more background correction that we need to perform. In
our deepest observation, 3419, the diffuse spectrum above 2 keV is fully
consistent with zero. In 3013 however, there is an obvious hard excess that
cannot originate from the sky because of its unphysical spectrum. This is
the excess discussed in \S\ref{sec:lc3013}, where we proposed that it is a
long flare of the same kind that affects the discarded time intervals in
this observation, only fainter.  Indeed, as the figure shows, the spectral
shape of the hard excess is consistent with the flare model derived in
\S\ref{sec:lc3013} (while being clearly inconsistent with an elevated
quiescent background, for example). The model normalization was fit to this
spectrum in the 2.5--10 keV interval.

Thus, we will proceed with the above assumption.  The flare model also gives
a significant contribution in the softer band. To correct it, we add this
best-fit model to the detector background spectrum (this correction was
already included in Fig.~\ref{fig:cumflux}).  Unfortunately, by doing so we
are setting the diffuse flux above 2 keV in field 3013 to zero by
definition, but such a correction is consistent with expectations from the
light curve and the flare spectrum. A normalization uncertainty of this
flare model ($\pm14$\%, $1\sigma$) will be included in the error budget for
this observation.

Figure~\ref{fig:residflare} also shows marginally significant excesses above
2 keV in OBSIDs 869 and 930, which can also be described by the flare model
(for those fits, we adopted the parameters derived for the 930 flare in
\S\ref{sec:lc3013}). However, while the excess spectrum in 3013 does match
the flare model, the same cannot be said with certainty for these two
observations. The excesses are also comparable to the quiescent background
uncertainty. As the figure shows, the possible flare contamination below 1
keV in these observations is negligible and can be safely ignored in the
analysis below. However, the above discussion illustrates the difficulty of
the diffuse CXB measurements at $E\gax 2$ keV using the ACIS BI chip S3; FI
chips may be better suited for that band (\S\ref{sec:future}).

In Appendix, we summarize the above background subtraction procedure and
describe how it can be applied for other ACIS observations of extended
sources.

\section{RESULTS}

Table 2 gives total and diffuse fluxes in our four CXB
fields in three energy bands. It includes both the directly measured
unresolved fluxes (that is, excluding the detected sources) and the
estimated true diffuse fluxes after the small correction for undetected
sources (\S\ref{sec:src} and Fig.~\ref{fig:cumflux}). The 2--7 keV part of
the table omits OBSID 3013 and the diffuse fluxes for OBSIDs 869 and 930
because of the residual background flare uncertainty (\S\ref{sec:3013resid})
that affects this band most significantly. The errors on these fluxes
include statistical errors of the data and the Moon background, and
systematic uncertainties of the background.  For ease of comparison with
earlier results, Table 2 also gives the normalization of a
power-law fit to the total spectra in the 1--7 keV band, with photon index
fixed at $-1.4$ (a slope consistent with all four spectra, although only
marginally for field 930 because of a strong soft excess), assuming Galactic
absorption. The average of the four fields is $10.7\pm0.9$ \phcmskevsr. No
attempt was made to correct our ``total'' fluxes for the missing very bright
sources that we were unlikely to encounter due to the small field of view
(covering a total of 0.07 deg$^2$), so these fluxes may not be
representative of the sky average. The brightest sources found in our fields
have unabsorbed 0.5--2 keV fluxes between $1-5\times 10^{-14}$ \ergscm.

\vspace{4mm}
\noindent
%%%%%%%%%%%%%%%%%%%%%%%%%%%%%%%%%%%%%%%%%%%%%%%%%%%%%%%%%%%%%%%%%%%%%%%%%%
%\begin{table*}[b]
%\begin{center}
\begin{minipage}{8.75cm}
\renewcommand{\arraystretch}{1.3}
\renewcommand{\tabcolsep}{1.8mm}
\small
\begin{center}
%
%\caption{Wide-Band Fluxes}
\centerline{\sc Table 2}
\vspace{1mm}
\centerline{\sc Wide-Band Fluxes}
\vspace{3mm}
\begin{tabular}{p{2cm}cccc}
\hline 
\hline
OBSID \dotfill         & 3013       & 3419        & 869        & 930 \\
\hline
0.3--1 keV$^a$ &&&\\
total   \dotfill       &$4.3\pm0.2$ &$2.2\pm0.2$  &$7.0\pm0.2$ &$7.8\pm0.2$ \\
unresolved \dotfill    &$2.7\pm0.2$ &$1.6\pm0.2$  &$6.7\pm0.2$ &$7.4\pm0.2$ \\
diffuse$^b$ \dotfill   &$2.4\pm0.2$ &$1.4\pm0.2$  &$6.4\pm0.2$ &$7.2\pm0.2$ \\
\hline
1--2 keV$^a$ &&&\\
total   \dotfill       &$1.1\pm0.2$ &$1.3\pm0.2$  &$1.0\pm0.2$ &$1.5\pm0.2$ \\
unresolved \dotfill    &$0.3\pm0.2$ &$0.6\pm0.15$ &$0.7\pm0.2$ &$1.2\pm0.2$ \\
diffuse$^b$ \dotfill   &$0.1\pm0.2$ &$0.3\pm0.15$ &$0.4\pm0.2$ &$1.0\pm0.2$ \\
\hline
2--7 keV$^a$ &&&\\
total   \dotfill       & ...       &$4.0\pm1.5$   &$4.7\pm1.8$ &$4.5\pm1.5$ \\
unresolved \dotfill    & ...       &$1.6\pm1.6$   & ...        & ...       \\
diffuse$^b$ \dotfill   & ...       &$0.8\pm1.6$   & ...        & ...       \\
\hline
Power law$^c$ &&&\\
total \dotfill         &$8.9\pm1.6$&$12.1\pm1.9$  &$8.8\pm1.6$ &$12.8\pm1.7$\\
\hline
\label{table:fluxes}
\end{tabular}
\end{center}
\vspace{-6mm}
{\footnotesize
${^a}$ In units of $10^{-15}$ \ergscmam, uncorrected for absorption.\\
${^b}$ Unresolved flux minus estimated contribution of undetected
sources.\\
${^c}$ Normalization of a power law fit to the 1--7 keV total spectrum,
corrected for absorption.
Photon index is fixed at 1.4, units are \phcmskevsr\ at $E=1$ keV.
}
\end{minipage}
%\end{center}
%\end{table*}
%%%%%%%%%%%%%%%%%%%%%%%%%%%%%%%%%%%%%%%%%%%%%%%%%%%%%%%%%%%%%%%%%%%%%%%%%%
\vspace{3.5mm}

\begin{figure*}[t]
\pspicture(0,14.8)(18.5,24)
%\psgrid(0,0)(18.5,24)

\rput[tl]{0}(1.,24){\epsfxsize=16cm \epsfclipon
\epsffile[18 204 588 500]{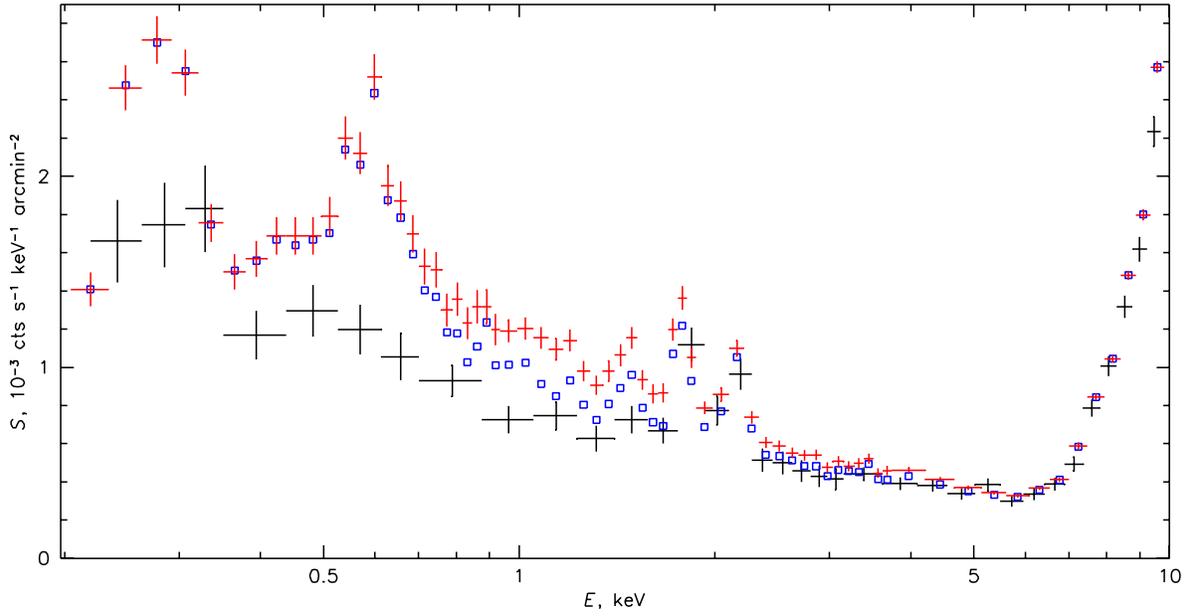}}

\rput[tl]{0}(0.0,15.6){
\begin{minipage}{18.5cm}
\small\parindent=3.5mm

\caption{Spectra of OBSID 3419 (our field with the lowest diffuse signal)
  and the Moon background. Black crosses show the Moon renormalized to match
  the 10--12 keV rate of 3419 (see \S\ref{sec:timedep}). Red crosses show
  the total spectrum (excluding the target) and blue squares show the
  diffuse component after the exclusion of all point sources (blue error
  bars are similar to the red ones and are omitted for clarity).}
\label{fig:moontotdiff}
\par
\end{minipage}
}

\endpspicture
\end{figure*}

\begin{figure*}[t]
\center
\pspicture(0,13.6)(9,23.4)
%\psgrid(0,0)(18.5,24)

\rput[tl]{0}(-0.2,24){\epsfxsize=9cm \epsfclipon
\epsffile{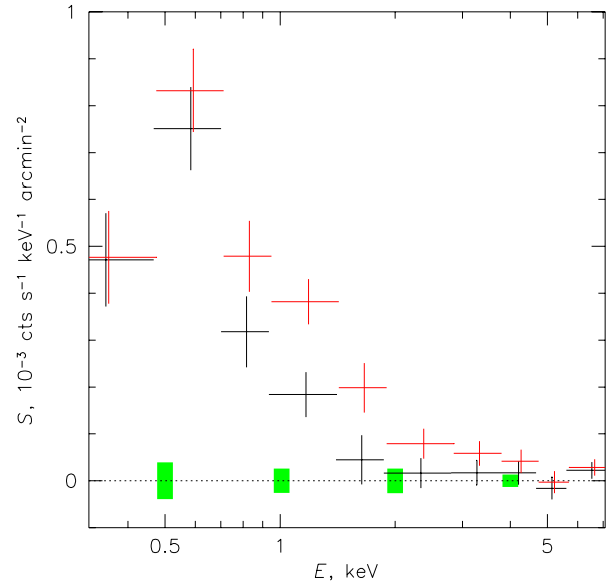}}

\rput[tl]{0}(0.0,15.4){
\begin{minipage}{8.75cm}
\small\parindent=3.5mm

\caption{Spectra of OBSID 3419 after the detector background
  subtraction. Red shows the total spectrum and black shows the diffuse
  component from Fig. \ref{fig:moontotdiff}.  Errors are dominated by the
  low statistics of the Moon. Green bars illustrate a 90\% systematic
  uncertainty of the quiescent background normalization of $\pm 3$\%. Above
  2 keV, the diffuse component is consistent with zero.}
\label{fig:totdiff}
\par
\end{minipage}
}

\endpspicture
\end{figure*}

Figure \ref{fig:moontotdiff} shows relative contributions of different
background components --- instrumental, cosmic diffuse, and point sources
--- in our cleanest observation, 3419, which also has the lowest soft
diffuse signal. Figure~\ref{fig:totdiff} shows the same spectra of the
diffuse and total (including point sources) CXB after the instrumental
background subtraction. Below 1 keV, a bright diffuse component dominates
the CXB spectrum. In the 1--2 keV band, point sources dominate, continuing
into the 2--5 keV band where the diffuse component disappears.

\begin{figure*}[t]
\pspicture(0,5.4)(18.5,23.4)
%\psgrid(0,0)(18.5,24)

\rput[tl]{0}(0.5,24){\epsfxsize=9.0cm \epsfclipon
\epsffile{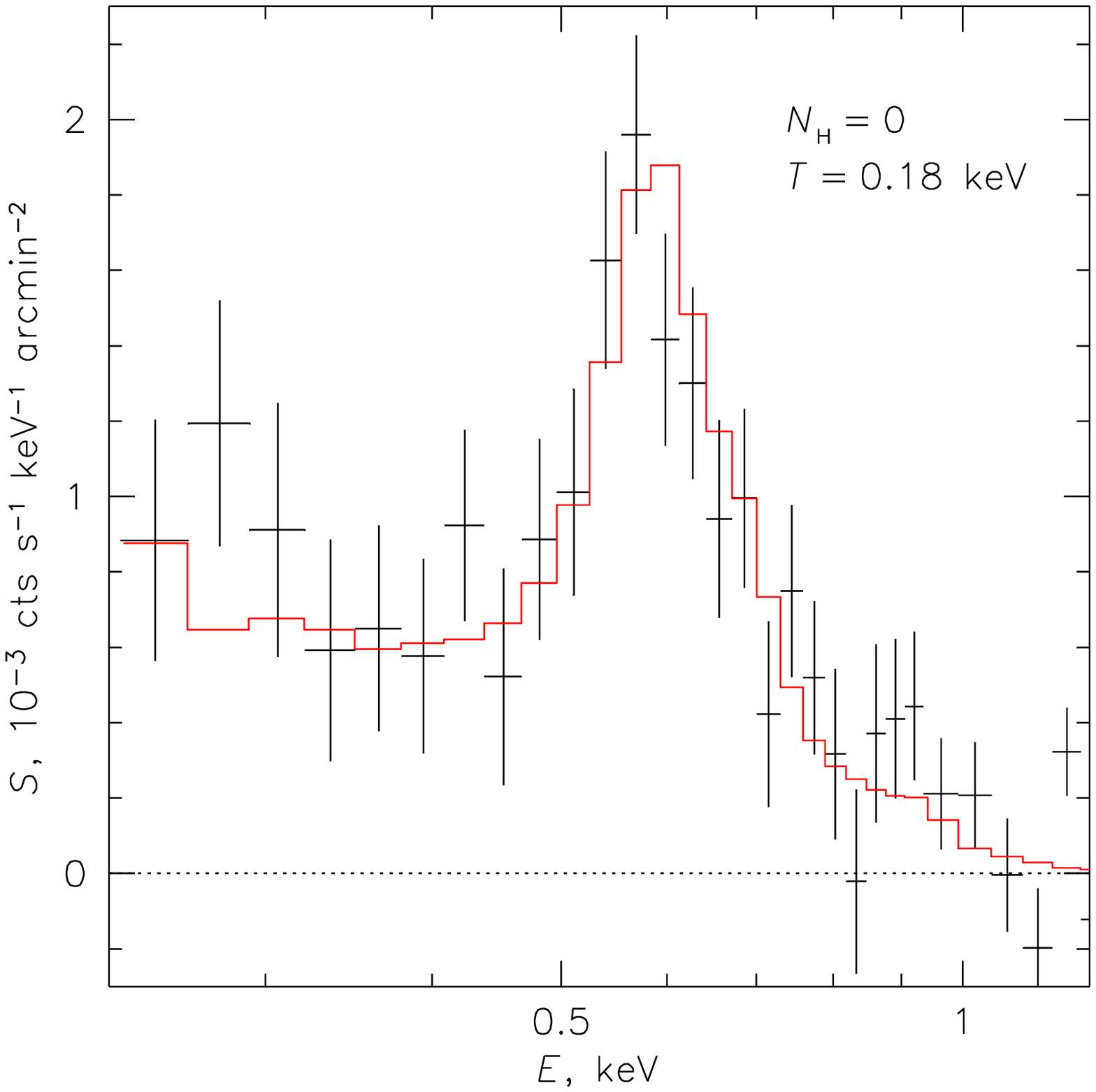}}

\rput[tl]{0}(9.5,24){\epsfxsize=9.0cm \epsfclipon
\epsffile{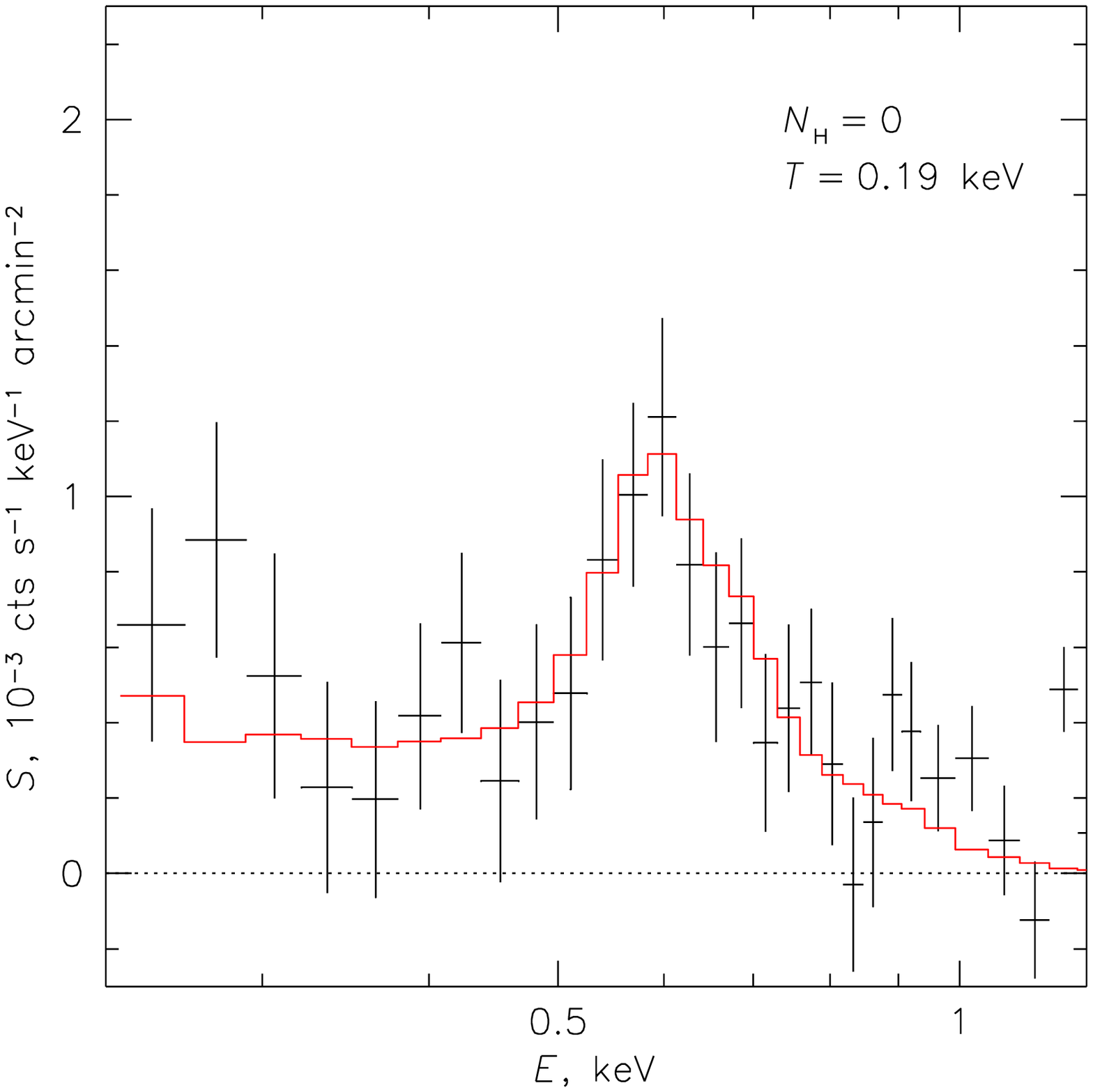}}

\rput[tl]{0}(0.5,15.8){\epsfxsize=9.0cm \epsfclipon
\epsffile{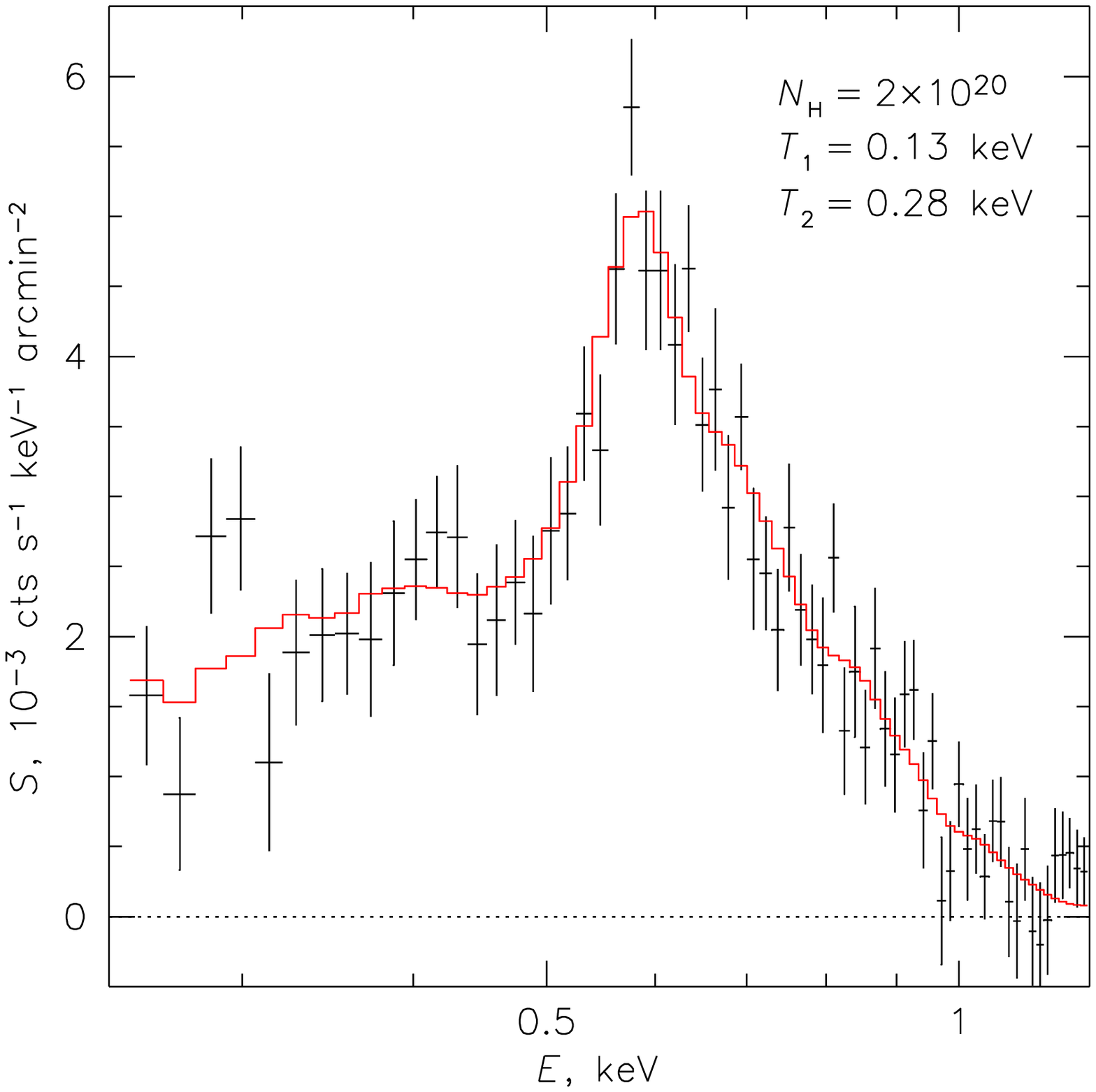}}

\rput[tl]{0}(9.5,15.8){\epsfxsize=9.0cm \epsfclipon
\epsffile{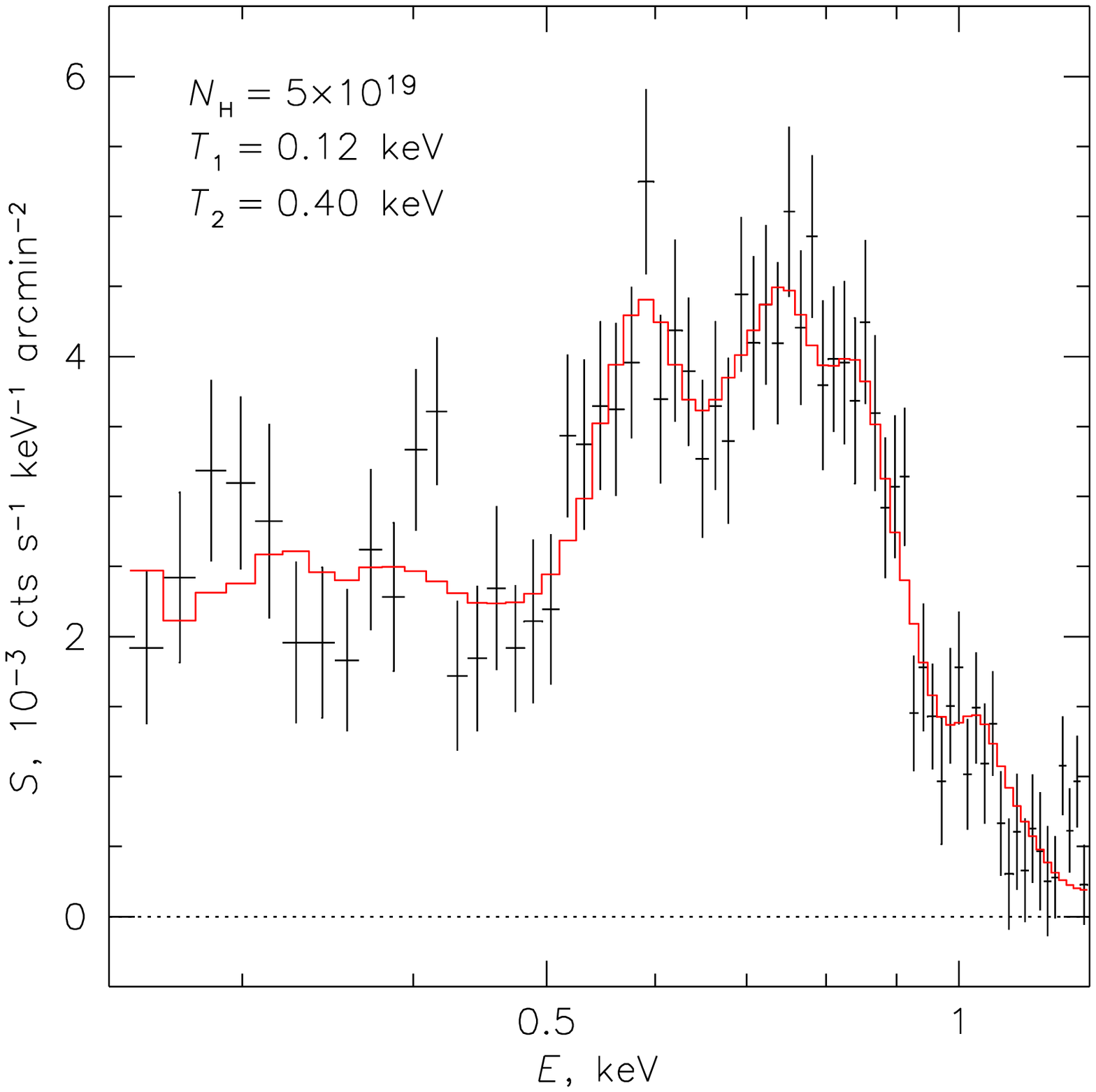}}

\rput[bl]{0}(2.3,22.6){\bf 3013}
\rput[bl]{0}(11.3,22.6){\bf 3419}
\rput[bl]{0}(2.3,14.4){\bf 869}
\rput[bl]{0}(16.8,14.4){\bf 930}

\rput[tl]{0}(0.0,7.0){
\begin{minipage}{18.5cm}
\small\parindent=3.5mm

\caption{The final soft diffuse spectra. OBSIDs are marked in each panel.
  For 3013, the residual flare is subtracted (see
  Fig.~\ref{fig:residflare}). Histograms show simple one- or two-temperature
  thermal plasma fits with parameters shown in the panels.}
\label{fig:softspec}
\par
\end{minipage}
}

\endpspicture
\end{figure*}

\begin{figure*}[t]
\center
\pspicture(0,13.6)(9,23.4)
%\psgrid(0,0)(18.5,24)

\rput[tl]{0}(-0.2,24){\epsfxsize=9cm \epsfclipon
\epsffile{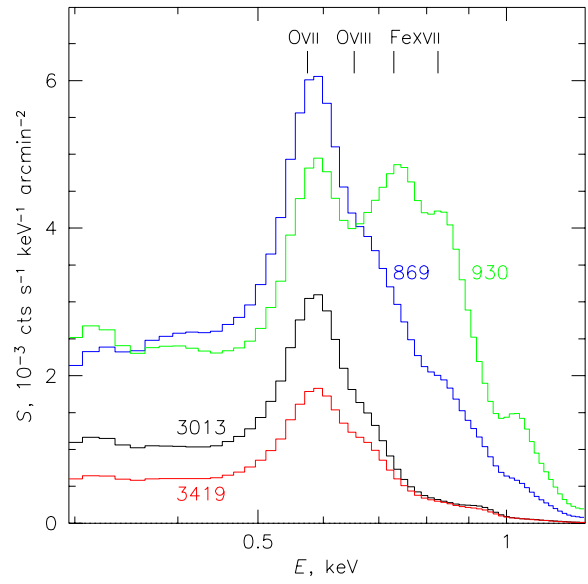}}

\rput[tl]{0}(0.0,15.4){
\begin{minipage}{8.75cm}
\small\parindent=3.5mm

\caption{Model fits for different fields from Fig.~\ref{fig:softspec}. To be
  directly comparable, models are convolved with the same telescope response
  without taking into account its time change.  OBSIDs and major line
  energies are marked.}
\label{fig:softmod}
\par
\end{minipage}
}

\endpspicture
\end{figure*}

Figure \ref{fig:softspec} shows the final diffuse spectra in the 0.25--1.2
keV band, with field 3013 corrected for the residual flare
(\S\ref{sec:3013resid}). All exhibit a ubiquitous O{\small VII} feature
around 570 eV (the He-like K$\alpha$ blend), with possible contributions
from K$\beta$ at 665 eV and the O{\small VIII} Ly$\alpha$ line at 654 eV
(these components cannot be resolved with the present statistics). The
best-fit mean energies and fluxes of this line blend are given in
Table~3. The line brightness strongly increases in the
direction of the North Polar Spur (fields 869 and 930), and an additional
pair of bright features (around 730 eV and 820 eV, primarily from Fe{\small
XVII}) emerges from the brighter part of the Spur (field 930). The lines are
fit well with a simple thermal plasma model (MEKAL, Kaastra 1992) with solar
abundances, consisting of either one component (for the two fields off the
Spur) or a two-temperature mixture (the Spur fields). The model parameters
are given in Fig.~\ref{fig:softspec}; the temperature range is 0.1--0.4 keV
and the best-fit absorption column is lower than the full Galactic value for
all four observations. For ease of comparison, the best-fit models from
Fig.~\ref{fig:softspec} are plotted together in Fig.~\ref{fig:softmod},
taking away the time dependence of the ACIS efficiency.

There is also an apparent line-like feature at $E\sim 300$ eV seen in all
four spectra. Its brightness appears to change together with that of the O
lines, and it is not present in the Moon spectrum
(Fig.~\ref{fig:moontotdiff}), which suggests that it is real. However,
because of the present calibration uncertainties at the lowest energies, it
is difficult to quantify its flux and even assess its reality. It should
also be kept in mind that the same Moon spectrum with large statistical
uncertainties was used as the background for all four observations.

\vspace{4mm}
\noindent
%%%%%%%%%%%%%%%%%%%%%%%%%%%%%%%%%%%%%%%%%%%%%%%%%%%%%%%%%%%%%%%%%%%%%%%%%%
%\begin{table*}[t]
%\begin{center}
\begin{minipage}{8.75cm}
\renewcommand{\arraystretch}{1.3}
\renewcommand{\tabcolsep}{1.5mm}
\small
\begin{center}
%
%\caption{Oxygen Line and Continuum Fluxes}
\centerline{\sc Table 3}
\centerline{\sc Oxygen Line and Continuum Fluxes}
\vspace{-1mm}
\begin{tabular}{p{3.1cm}cccc}
\hline 
\hline
OBSID \dotfill & 3013           & 3419       & 869       & 930 \\
\hline
O line energy, eV \dotfill %
               &$570\pm10$&     $580\pm14$ & $590\pm 6$   &$585\pm8$\\
O line flux$^a$ \dotfill %
               &$9.8\pm1.5$ &   $5.6\pm1.5$ &$15\pm 2$ & $12\pm 2$\\
0.3--1 keV continuum$^b$\dotfill %
               &$3.2\pm 0.5$&   $3.5\pm1.0$  & ...       & ...\\
\hline
\label{table:ofluxes}
\end{tabular}
\end{center}
\vspace{-6mm}
{\footnotesize
${^a}$ In units of $10^{-7}$ \phscmam, uncorrected for absorption.\\
${^b}$ Flux excluding O line, in units of $10^{-15}$ \ergscmam,
corrected for absorption assuming full Galactic column.
}
\end{minipage}
%\end{center}
%\end{table*}
%%%%%%%%%%%%%%%%%%%%%%%%%%%%%%%%%%%%%%%%%%%%%%%%%%%%%%%%%%%%%%%%%%%%%%%%%%
\vspace{3mm}

\section{DISCUSSION}

\subsection{Comparison with earlier results}

We can compare our total (diffuse + sources) CXB results with the recent
\xmm\ work (Lumb et al.\ 2002). Since the brightest sources in our four
fields have fluxes similar to the lowest fluxes of the excluded sources in
Lumb et al.\ ($1-2\times 10^{-14}$ \ergscm\ in the 0.5--2 keV band), our
``total'' fluxes can be directly compared to their values after the bright
source exclusion. Starting from high energies, normalizations of our
power-law fits in the 1--7 keV band given in Table 2 are in
agreement with the \xmm\ average of 8.4 \phcmskevsr\ at 1 keV, if we exclude
our bright field 930.  In the narrower 1--2 keV band, our fluxes (again,
except 930) are in good agreement with Lumb et al.'s $1.0\times 10^{-15}$
\ergscmam\ calculated from their best-fit model. Our 2--7 keV fluxes also
agree with $3.2\times 10^{-15}$ \ergscmam\ converted from their 2--10 keV
flux (we note here that these three values in Table 2 are
not entirely statistically independent, because the error is dominated by
the same Moon dataset). Our uncertainties in the 2--7 keV band are large;
future ways to reduce them are described in \S\ref{sec:future} below.

Our 1--7 keV power law normalizations also agree with $9-11$ \phcmskevsr\ at
1 keV derived by Miyaji et al.\ (1998) from \asca\ GIS data for two high
Galactic latitude fields, and with 11.7 \phcmskevsr\ derived by Vecchi et
al.\ (1999) using \sax.  Note that those studies cover much greater solid
angles and therefore are more likely to include rare, bright sources, so
this comparison is only approximate.

At $E<1$ keV, the scatter between our fields (a factor of 5) is higher than
that in Lumb et al.\ (a factor of 2--3), but this, of course, is because of
our specific selection of fields spanning a range of \rosat\ fluxes.

At the primary O line energy, the CXB is dominated by the diffuse component
(Fig.~\ref{fig:totdiff}). Thus, we can compare our line fluxes to those
recently derived in a microcalorimetric experiment by McCammon et al.\
(2002) from a 1 sr area mostly away from bright Galactic features. They
report an average flux in the O{\small VII} + O{\small VIII} lines of
$(5.4\pm 0.8) \times 10^{-7}$ \phscmam. This is in the range of our values
for the off-Spur observations 3013 and 3419 given in Table 3.  McCammon et
al.\ (2002) also observed lines at lower energies, some of which may explain
our 300 eV feature (if it is real).  Oxygen line fluxes derived from earlier
experiments (e.g., Inoue et al.\ 1979; Gendreau et al.\ 1995) are also
within our range. Our results show, however, that even for these high
Galactic latitude areas away from bright Galactic features, the line
brightness varies significantly from field to field. The general shape of
the North Polar Spur spectrum in our field 930 is consistent with that
reported in earlier works (e.g., Schnopper et al.\ 1982; Rocchia et al.\
1984; Warwick 2002).

\subsection{Extragalactic diffuse component}

The diffuse flux in fields 869 and 930 is obviously dominated by the North
Polar Spur. The O blend in 3013 and 3419 probably has a Galactic or local
origin as well (extragalactic sources with redshifts $>0.05$ are excluded by
our measured line energy; furthermore, the high-resolution spectrum of
McCammon et al.\ 2002 excludes {\em any}\/ redshift), although we cannot
exclude, for example, their Local Group origin. Thus, the continuum
component in the two low-brightness fields can give an approximate upper
limit on the flux from the vast quantities of the putative warm
intergalactic gas (e.g., Cen \& Ostriker 1999) that should emit a mixture of
lines and continua from different redshifts. For a conservatively high
estimate of this continuum component, we fit the 0.3--1 keV diffuse spectra
by a power law model plus the line, applying the full Galactic absorbing
column.

The resulting unabsorbed continuum fluxes are given in Table 3; they
correspond to the spectral density of $(2.4-2.5)\times 10^{-15}$
\ergscmam\,keV$^{-1}$ at $E=0.7$ keV. This is well above the typical
theoretical predictions that range between $(0.3-1)\times 10^{-15}$
\ergscmam\,keV$^{-1}$ (e.g., Cen \& Ostriker 1999; Phillips, Ostriker, \&
Cen 2001; Voit \& Bryan 2001; but see Bryan \& Voit 2001 for a higher
predicted flux from simulations without the inclusion of cooling and
preheating). However, the predicted average brightness values from the
published simulations are not directly comparable to our result, because
they are dominated by nearby galaxy groups and clusters that are easily
detected and excluded from our and other CXB measurements.

Thus, our crude estimate does not constrain the warm intergalactic gas
models. The constraint may be improved in the future by better modeling and
subtraction of the Galactic emission and spatial fluctuation analysis (as
in, e.g., Kuntz, Snowden \& Mushotzky 2001); however, since the Galaxy
dominates at these energies, such constraints will necessarily be
model-dependent.

\subsection{Origin of line emission}

The observed energies of the spectral lines suggest local (in the Local
Group, Galaxy, or our immediate vicinity) origin of the dominant fraction of
the soft diffuse CXB. Extensive literature exists that models its various
components under the well-justified assumption of their thermal plasma
origin (e.g., Kuntz \& Snowden 2000 and references therein). Leaving such
modeling for future work, here we mention an interesting alternative
possibility.

It is likely that a significant fraction of the line flux comes from charge
exchange (CX) between highly charged ions in the solar wind (primarily bare
and hydrogenic O and C) and neutral gas occurring throughout the heliosphere
and in the geocorona. Most of the flux that would be observed from
heliospheric CX (with H and He) originates within a few tens of AU of the
Sun. Geocoronal emission arises where residual atmospheric H is exposed to
the solar wind, at distances of order 10 Earth radii.

In the CX process, a collision between a solar wind ion and a neutral atom
leads to the transfer of an electron from the neutral species to a high-$n$
energy level in the ion, which then decays and emits an X ray. Dennerl et
al.\ (1997) and Cox (1998) were the first to suggest that these photons
might contribute to the CXB, and Cravens (2000) estimated that heliospheric
emission might account for roughly half of the observed soft CXB. Cravens,
Robertson, \& Snowden (2001) also argued that the excess time-variable
diffuse flux often observed by \rosat\ was due to fluctuations in
heliospheric and especially geocoronal CX emission, caused by ``gusts'' in
the solar wind.

In the \chandra\ Moon observations, the heliospheric component will be
blocked, but geocoronal emission should be present. We estimate, however,
that the typical intensity of that signal, about a few $\times 10^{-8}$ phot
s$^{-1}$ cm$^{-2}$ arcmin$^{-2}$ in O K$\alpha$ (the strongest line), is
smaller than the statistical uncertainties in our measurement.  Note also
that the geocoronal signature is not likely to be present in our net CXB
spectra, since it is subtracted as part of the Moon spectrum.  Heliospheric
CX emission should be stronger and less time-variable than geocoronal
emission, and should be present in our spectra. We have constructed a
numerical model, based on Cravens' (2000) work and similar to that described
in Wargelin \& Drake (2001, 2002), that predicts that roughly half the flux
in the O line(s) in fields 3013 and 3419 may come from heliospheric
CX. Those observations were at low ecliptic latitude, within the ``slow''
and more highly ionized solar wind. CX flux in fields 869 and 930 is
expected to be much lower because those observations looked through the
``fast'' solar wind, which has a much smaller fraction of bare and H-like O
ions (von Steiger et al.\ 2000). Observations of a sample of fields selected
specifically to test this possibility are required for a more quantitative
analysis, which is forthcoming (B. J. Wargelin et al.\ in preparation).

\subsection{Future work}
\label{sec:future}

For further CXB studies with \chandra, it is useful to look into the error
budget of the present results. The errors in the 1--7 keV band are dominated
by the detector background uncertainty: the statistical error of the short
Moon dataset, the uncertainty on its normalization, and the possible
residual flare component.  Forthcoming calibration observations with ACIS
stowed but working in the full imaging mode should take care of the first
component. The other two point toward the use of FI chips for studies in
this band. The scatter of the quiescent background normalization probably
cannot be reduced below our 2--3\% estimate; however, the FI detector
background itself is lower by a factor of 2--3, depending on the energy. The
FI chips also are much less affected by the background flares. An ACIS-I
study of the CXB will be presented in a forthcoming paper (S. Virani et al.,
in preparation).

\section{SUMMARY}

We have analyzed four high Galactic latitude, empty fields observed with
\chandra\ ACIS-S3 and, for the first time, derived spectra of the diffuse
X-ray background, directly excluding the point source contribution. The
total (diffuse and point sources) CXB brightness in all bands is in
agreement with most previous experiments.  In the 0.3--1 keV band, the
diffuse signal varies strongly between all four fields. In the two fields
far from known bright Galactic features, it contributes about half of the
total CXB seen by instruments with poorer angular resolution.  It is
dominated by emission lines (most prominently, the O{\small VII} + O{\small
VIII} blend at $E\approx 580$ eV, indications of which were also seen in
previous experiments) and can be described by a thermal plasma model with
$kT=0.1-0.2$ keV. The line brightness increases strongly, and additional
lines appear, in the directions of the North Polar Spur.

At higher energies, the background is more uniform. Diffuse emission is
detected with high significance in the 1--2 keV band in the brighter of the
two North Polar Spur fields. In other fields at $E>1$ keV, the diffuse
component is weak or consistent with zero at our present accuracy (to be
improved in the forthcoming analysis of ACIS-I deep observations). Our
current uncertainties for the diffuse CXB --- one of the lowest surface
brightness celestial objects --- are large due to the difficulty of modeling
the instrumental background in the time-variable particle environment
encountered in high-altitude orbits like \chandra's.

\acknowledgements

The results presented here are made possible by the successful effort of the
entire \chandra\ team to launch and operate the observatory.  We thank the
referee for helpful comments and suggestions.  Partial support was provided
by NASA contract NAS8-39073, grant NAG5-9217, and the Smithsonian
Institution.

\begin{appendix}

\section{ACIS BACKGROUND SUBTRACTION}

For the ACIS analysis of extended objects, the background can be modeled and
subtracted as we did in \S\ref{sec:analysis}. In addition to the
July-September 2001 dark Moon data and the observations with ACIS stowed
(forthcoming in 2002 and 2003) that contain only the detector background,
large datasets combining a number of high Galactic latitude, relatively
source-free fields are available (see Markevitch 2001 for details). The
latter data are cleaned of flares and point sources, but include the cosmic
diffuse emission, so they are relevant when this emission is part of the
background as opposed to signal.  Separate blank-sky datasets are assembled
for three ACIS temperatures (which determines the spectral resolution), and
for different time periods for the present $-120^\circ$C temperature, to
track the background evolution.

The following steps summarize our background subtraction procedure as it
could be applied in a typical extended source analysis, using the combined
blank field datasets. In anomalous cases, this procedure may need more than
one iteration.

1. To identify flares, we select a region of the chip(s) free of bright
source emission and use it to create a light curve with time bins large
enough to be able to detect rate changes by a factor of 1.2 (the fiducial
factor used in creation of the blank sky datasets).  Using the relevant
background dataset (determined by the observation date), we evaluate the
nominal count rate in the selected region, and reject all intervals with the
rate above a factor of 1.2 of this expected rate (usually, the flares are
well-defined in time and the exact threshold is not important). BI and FI
chips should be cleaned separately because BI chips are much more prone to
flares. Given the spectra of the flares, for the BI chips, a 2.5--7 keV (or
2.5--6 keV for S1) energy band should be used, while for the FI chips, the
full 0.3--12 keV band may be used.

If the target is in S3 and covers the whole chip, chip S1 can be used for
flare detection. Despite the different quiescent backgrounds, S1 and S3 show
a very similar response to flares, including the flare timing, spectra and
intensities.  If S1 is not available either, then faint, non-obvious flares
may escape detection and significantly affect the results, because flare and
quiescent background components have very different spectral shape (see,
e.g., a discussion in Markevitch 2002 and \S\ref{sec:lc3013} above).  In
such cases, one should at least check the validity of the results by trying
to fit a flare model to the spectrum above 2 keV extracted from
low-brightness target areas (as in \S\ref{sec:3013resid}). That spectral
model applies only to the ``soft'' flare species that does not affect the FI
chips.

If there is a faint, long flare in the observation whose proper exclusion
results in too little useful exposure left, and if there are source-free
areas in the chip of the same type, one can try to model the flare
contribution using those areas (as was attempted, e.g., in Markevitch et
al.\ 2002 and Markevitch 2002). However, the flare flux is not spatially
uniform; its spatial distribution is under investigation.

2. After the flare exclusion, the background dataset can be normalized by
the ratio of high-energy rates, e.g., in the PHA interval of 2500--3000
ADU. This ratio should be within $\pm10$\% of the exposure ratio. Item 1 may
need to be repeated with the corrected nominal rate if this correction is
large. A 90\% systematic uncertainty on the background normalization derived
in this manner is about 3\% (\S\ref{sec:timedep}); however, if any residual
faint flare is suspected, it should be taken into account in the final
uncertainty.

3. If the observation was performed in VF mode, and if a VF mode background
dataset is available for that time period, additional background filtering
(Vikhlinin 2001) can be applied.

4. As seen from our results (e.g., Table 2), the soft
($E\lax 1$ keV) CXB component varies strongly from field to field even at
high Galactic latitudes, so the blank-sky background may not have the
correct soft spectrum. One can, for example, compare the RASS R4-R5 flux
(Snowden et al.\ 1997) for a given observation to the average R4-R5 flux for
the background dataset to determine if a correction is needed (for low-$b$
fields, it may be needed even if those wide-band fluxes are similar, since
the soft spectra may still be very different). If so, one can try to use
regions of the field of view free of sources (not necessarily in the same
chip) to model the sky soft excess or deficiency w.r.t.\ the blank-sky
background (e.g., by the simple models used in this paper), and subtract
this model from the regions of interest after the proper vignetting
correction (as, e.g., in Markevitch \& Vikhlinin 2001 and Markevitch
2002). An alternative is to use the dark Moon data or the ACIS-stowed data
(if the observation date is sufficiently close to those datasets) and model
away the whole diffuse CXB component.

\end{appendix}

%%%%%%%%%%%%%%%%%%%%%%%%%%%%%%%%%%%%%%%%%%%%%%%%%%%%%%%%%%%%%%%%%%%%%%%%%%
%%%%%%%%%%%%%%%%%%%%%%%%%%%%%%%%%%%%%%%%%%%%%%%%%%%%%%%%%%%%%%%%%%%%%%%%%%

%%%%%%%%%%%%%%%%%%%%%%%%%%%%%%%%%%%%%%%%%%%%%%%%%%%%%%%%%%%%%%%%%%%%%%%%%%
%%%%%%%%%%%%%%%%%%%%%%%%%%%%%%%%%%%%%%%%%%%%%%%%%%%%%%%%%%%%%%%%%%%%%%%%%%

%%%%%%%%%%%%%%%%%%%%%%%%%%%%%%%%%%%%%%%%%%%%%%%%%%%%%%%%%%%%%%%%%%%%%%%%%%
%%%%%%%%%%%%%%%%%%%%%%%%%%%%%%%%%%%%%%%%%%%%%%%%%%%%%%%%%%%%%%%%%%%%%%%%%%

%%%%%%%%%%%%%%%%%%%%%%%%%%%%%%%%%%%%%%%%%%%%%%%%%%%%%%%%%%%%%%%%%%%%%%%%%%
%%%%%%%%%%%%%%%%%%%%%%%%%%%%%%%%%%%%%%%%%%%%%%%%%%%%%%%%%%%%%%%%%%%%%%%%%%

%%%%%%%%%%%%%%%%%%%%%%%%%%%%%%%%%%%%%%%%%%%%%%%%%%%%%%%%%%%%%%%%%%%%%%%%%%
%%%%%%%%%%%%%%%%%%%%%%%%%%%%%%%%%%%%%%%%%%%%%%%%%%%%%%%%%%%%%%%%%%%%%%%%%%

%%%%%%%%%%%%%%%%%%%%%%%%%%%%%%%%%%%%%%%%%%%%%%%%%%%%%%%%%%%%%%%%%%%%%%%%%%
%%%%%%%%%%%%%%%%%%%%%%%%%%%%%%%%%%%%%%%%%%%%%%%%%%%%%%%%%%%%%%%%%%%%%%%%%%

%%%%%%%%%%%%%%%%%%%%%%%%%%%%%%%%%%%%%%%%%%%%%%%%%%%%%%%%%%%%%%%%%%%%%%%%%%
%%%%%%%%%%%%%%%%%%%%%%%%%%%%%%%%%%%%%%%%%%%%%%%%%%%%%%%%%%%%%%%%%%%%%%%%%%

%%%%%%%%%%%%%%%%%%%%%%%%%%%%%%%%%%%%%%%%%%%%%%%%%%%%%%%%%%%%%%%%%%%%%%%%%%
%%%%%%%%%%%%%%%%%%%%%%%%%%%%%%%%%%%%%%%%%%%%%%%%%%%%%%%%%%%%%%%%%%%%%%%%%%

%%%%%%%%%%%%%%%%%%%%%%%%%%%%%%%%%%%%%%%%%%%%%%%%%%%%%%%%%%%%%%%%%%%%%%%%%%
%%%%%%%%%%%%%%%%%%%%%%%%%%%%%%%%%%%%%%%%%%%%%%%%%%%%%%%%%%%%%%%%%%%%%%%%%%

%%%%%%%%%%%%%%%%%%%%%%%%%%%%%%%%%%%%%%%%%%%%%%%%%%%%%%%%%%%%%%%%%%%%%%%%%%
%%%%%%%%%%%%%%%%%%%%%%%%%%%%%%%%%%%%%%%%%%%%%%%%%%%%%%%%%%%%%%%%%%%%%%%%%%

%%%%%%%%%%%%%%%%%%%%%%%%%%%%%%%%%%%%%%%%%%%%%%%%%%%%%%%%%%%%%%%%%%%%%%%%%%
%%%%%%%%%%%%%%%%%%%%%%%%%%%%%%%%%%%%%%%%%%%%%%%%%%%%%%%%%%%%%%%%%%%%%%%%%%

%%%%%%%%%%%%%%%%%%%%%%%%%%%%%%%%%%%%%%%%%%%%%%%%%%%%%%%%%%%%%%%%%%%%%%%%%%
%%%%%%%%%%%%%%%%%%%%%%%%%%%%%%%%%%%%%%%%%%%%%%%%%%%%%%%%%%%%%%%%%%%%%%%%%%

\end{document}